\documentclass[twocolumn]{aastex61}

\usepackage{txfonts}
\usepackage{xspace}
\usepackage{todonotes}

\newcommand{\code}[1]{\texttt{#1}}

\newcommand{\mesa}{\code{MESA}}

\newcommand{\pdcz}{pulse-driven convective zone}

\newcommand{\spr}{\mbox{$s$-process}}
\newcommand{\sprn}{\mbox{$s$ process}}

\newcommand{\Mzams}{\ensuremath{\unitstyle{M}_{ZAMS}}}

\newcommand{\numberspace}{\ensuremath{\;}}

\newcommand{\unitstyle}[1]{\ensuremath{\mathrm{#1}}}

\newcommand{\Msun}{\ensuremath{\unitstyle{M}_\odot}}

\newcommand{\unit}[2]{\ensuremath{#1\numberspace\mathrm{#2}}}

\newcommand{\lSect}[1]{{\label{sec:#1}}}
\newcommand{\lFig}[1]{{\label{fig:#1}}}

\newcommand{\pan}[1]{{\textit{#1}}}

\newcommand{\FIGFF}[2]{{\ref{fig:#2}\pan{#1}}}

\newcommand{\FIG}[2]{{Fig.\,\FIGFF{#1}{#2}}}
\newcommand{\Fig}[1]{{\FIG{}{#1}}}

\newcommand{\Sectff}[1]{{\ref{sec:#1}}}
\newcommand{\Sect}[1]{{\S\Sectff{#1}}}

\newcommand{\isofont}[1]{{\mathrm{#1}}}
\newcommand{\isomass}[1]{{\ensuremath{\isofont{^{#1}}}}}
\newcommand{\isocharge}[1]{{\ensuremath{\isofont{_{#1}}}}}
\newcommand{\isotope}[3]{{\ensuremath{\isocharge{#1}\isomass{#2}\isofont{#3}}}}

\newlength{\apjcolwidth}
\newlength{\figwidth}
\newlength{\doublewide}
\newcommand{\cyberhs}[0]{\code{cyberhubs}} 
\newcommand{\cyberh}[0]{\code{cyberhub}} 
\newcommand{\neutron}{\ensuremath{n}}

\begin{document}
\title{Cyberhubs: Virtual Research Environments for Astronomy}
\author{Falk Herwig}
\affiliation{Department of Physics \& Astronomy, University of Victoria, Victoria, Canada}\affiliation{Joint Institute for Nuclear Astrophysics, Center for the Evolution of the Elements, Michigan State University, USA}
\affiliation{NuGrid Collaboration}

\author{Robert Andrassy}
\affiliation{Department of Physics \& Astronomy, University of Victoria, Victoria, Canada}
\affiliation{Joint Institute for Nuclear Astrophysics, Center for the Evolution of the Elements, Michigan State University, USA}

\author{Nic Annau}
\affiliation{Department of Physics \& Astronomy, University of Victoria, Victoria, Canada}

\author{Ondrea Clarkson}
\affiliation{Department of Physics \& Astronomy, University of Victoria, Victoria, Canada}
\affiliation{Joint Institute for Nuclear Astrophysics, Center for the Evolution of the Elements, Michigan State University, USA}
\affiliation{NuGrid Collaboration}

\author{Benoit C\^ot\'e}
\affiliation{Konkoly Observatory, Research Centre for Astronomy and Earth Sciences, Hungarian Academy of Sciences, Budapest, Hungary}
\affiliation{Department of Physics \& Astronomy, University of Victoria, Victoria, Canada}
\affiliation{Joint Institute for Nuclear Astrophysics, Center for the Evolution of the Elements, Michigan State University, USA}
\affiliation{NuGrid Collaboration}

\author{Aaron D'Sa}
\affiliation{LCSE, University of Minnesota, Minneapolis, USA}
\affiliation{Joint Institute for Nuclear Astrophysics, Center for the Evolution of the Elements, Michigan State University, USA}

\author{Sam Jones}
\affiliation{Los Alamos National Laboratory, Los Alamos, New Mexico, USA}
\affiliation{Joint Institute for Nuclear Astrophysics, Center for the Evolution of the Elements, Michigan State University, USA}
\affiliation{NuGrid Collaboration}

\author{Belaid Moa}
\affiliation{Research Computing Services, University of Victoria, Victoria, Canada}

\author{Jericho O'Connell}
\affiliation{Department of Physics \& Astronomy, University of Victoria, Victoria, Canada}

\author{David Porter}
\affiliation{MSI, University of Minnesota, Minneapolis, USA}

\author{Christian Ritter}
\affiliation{Department of Physics \& Astronomy, University of Victoria, Victoria, Canada}
\affiliation{Joint Institute for Nuclear Astrophysics, Center for the Evolution of the Elements, Michigan State University, USA}
\affiliation{NuGrid Collaboration}

\author{Paul Woodward}
\affiliation{LCSE, University of Minnesota, Minneapolis, USA}
\affiliation{Joint Institute for Nuclear Astrophysics, Center for the Evolution of the Elements, Michigan State University, USA}

\correspondingauthor{Falk Herwig}
\email{fherwig@uvic.ca}

\received{December 8, 2017}
\revised{Feb 1, 2018}
\accepted{acceptance date}
\published{published date}
\submitjournal{ApJS}


\begin{abstract}
Collaborations in astronomy and astrophysics are faced with numerous
cyber infrastructure challenges, such as large data sets, the need to
combine heterogeneous data sets, and the challenge to effectively
collaborate on those large, heterogeneous data sets with significant
processing requirements and complex science software tools. The
\cyberhs\ system is an easy-to-deploy package for small to medium-sized
collaborations based on the Jupyter and Docker technology, that allows
web-browser enabled, remote, interactive analytic access to shared
data. It offers an initial step to address these challenges. The
features and deployment steps of the system are described, as well as
the requirements collection through an account of the different
approaches to data structuring, handling and available analytic tools
for the NuGrid and PPMstar collaborations. NuGrid is an international
collaboration that creates stellar evolution and explosion physics and
nucleosynthesis simulation data. The PPMstar collaboration performs
large-scale 3D stellar hydrodynamics simulation of interior convection
in the late phases of stellar evolution. Examples of science that is
presently performed on \cyberhs, in the areas 3D stellar hydrodynamic
simulations, stellar evolution and nucleosynthesis and Galactic
chemical evolution, are presented.
\end{abstract}


\section{Introduction}
\lSect{intro} New astronomical observatories, large surveys, and the
latest generation of astrophysics simulation data sets, provide the
opportunity to advance our understanding of the universe
profoundly. However, the sheer size and complexity of the new data
sets dictate rethinking of the current data analytic practice which
can often be a barrier to fully exploiting the scientific potential of
these large data sets. The following challenges may be identified.

The typical \textbf{size of data sets in astronomy \& astrophysics}
continues to grow substantially. To name a few examples, optical
waveband projects such as the Large Synoptic Survey Telescope (LSST),
radio facilities such as the Square Kilometer Array (SKA) pathfinders,
as well as large data sets produced by cosmological or stellar
hydrodynamics simulations which will, in combination, produce 10s of
petabytes of scientific data that would not typically be held in one
place. Such data sets can not be downloaded for processing and
analysis. Instead combining remote computer processing capacity where
the data is stored with the appropriate analytic and data processing
software stacks is required. On top of that efficient remote access
with visualization and interaction capabilities are needed to enable a
distributed community to collectively explore these data sets. Instead
of moving the data to the processing machines, the processing
pipelines need to be moved to where the data resides.

One of the great promises of the age of data science for astronomy is
that the many multi-physics, multi-messenger, multi-wavelength and
multi-epoch constraints that bear on most problems in astronomy, can
in fact be combined successfully when complex data interactions and
collaborative cyber-research environments are constructed. This notion
of \textbf{data fusion} between different types of observational data
and simulation data can reach its full potential when different
communities can come together in combining and \textbf{sharing data
  sets, analytical tools and pipelines, and derived results}.

A significant barrier in accomplishing this goal in practice are
authentication and access models. Resources are often provided by
national bodies or consortia with often burdensome access requirements
and limitations. An effective research platform system would accept
easily-available social or otherwise broadly used third-party
identities to allow \textbf{flexible, international, collaborative
  access}. Although single sign-on technologies are emerging, the
scientific community is still far from adopting them.

Decades worth of investments in legacy software, tools, and workflows
by many research groups may be lost and never shared with a broader
community or applied to new data sets. The \textbf{reproducibility of
  science} is suffering when data processing and analytic workflows
can not be shared. A modern research platform would provide a uniform
execution environment that will \textbf{liberate legacy software},
unlocking its analytic value and that of associated data sets, making
them available to others, and allowing them to reproducibly interact
with analysis procedures and data sets of other researchers.

In this paper we describe \cyberhs, an easily-deployed service that
combines \code{Juypter} notebook or
\code{Jupyterlab} data and
processing in a containerized environment with a prescribed software
application toolbox and data collection.  The system described here is
the result of multi-year developments and evolution, with the goal to
address the limitations and challenges, initially of the international
NuGrid collaboration, and then in addition, those of the
PPMstar collaboration. Both of these brought somewhat different
demands and requirements, that in combination are likely typical for a
very wide range of use cases in astronomy and astrophysics.

The international Nuclesoynthesis Grid
(NuGrid\footnote{\url{http://www.nugridstars.og}})
collaboration has members from $\approx 20$ institutions in many
countries \citep{Pignatari:2012dw}. Since 2007, NuGrid has been
combining the required expertise from many different scientists to
generate the most comprehensive data sets for the production of the
elements in massive and low-mass stars to date
\citep{Pignatari:2016er,Ritter:2017wy}. Around 2009, when the first
data set was created it had a size of $\approx \unit{5}{TB}$, which is
relatively small by today's standards, however still large enough that
it is not easily transferred around the globe on demand. The
collaboration was faced with problems that are common to any
distributed, data-oriented collaboration. Problem-specific processing
and analytic tools are developed, but deployment in different places
is always complicated by a diverse set of computing environments. In
the NuGrid collaboration, which brings together different communities,
all of the three major operating systems are used. Initially, three
copies of the data sets were always maintained at three institutions
in the UK, Switzerland and Canada through a tedious, time-consuming
and error-prone syncing system. Still, analytic and interactive access
for researchers who were not at one of those three institutions was
very limited through VNC sessions or X11 forwarding, and required
account-level access sharing that would be probably impossible at most
institutions today.

As a next step, the collaboration adopted CANFAR's
\code{VOspace} as a
shared and mountable storage system. CANFAR is the Canadian Advanced
Network for Astronomy
Research\footnote{\url{http://www.canfar.net}}, a consortium
between NRC (National Research Council) Herzberg's Canadian Astronomy
Data Center (CADC) and Canadian university groups which aims to
jointly address astronomy cyber-infrastructure challenges. VOspace
provides shareable user storage with a web interface, and identity and
group management system, similar to commercial cloud storage systems
such as Google Drive and Dropbox. It also has a Python API
\code{vos} that
allows command line access to VOspace and POSIX mounting to a local
laptop or workstation. The mounting option, in particular, made
VOspace very promising for the collaboration, as it allows for the
replacement of three distributed storage copies with just one master
copy, which can be mounted from anywhere.  In addition, it includes a
smart indexing algorithm which ensures that only the data needed for
an analysis or plot is transferred. Although this system was a
significant improvement, it did not work for remotely executed
analysis projects requiring high data throughput of most of the
available data set, and it did not solve the problem that many in the
collaboration felt restricted by complications in establishing and
maintaining the NuGrid software stack. This software stack is not
particularly complex, from the viewpoint of a computationally
experienced user, but in order to address the diverse set of science
challenges, the collaboration includes members that deploy a diverse
set of science methodologies, and especially entry-level researchers
and researchers-in-training have often found establishing and
maintaining the NuGrid software stack to be a substantial barrier.

We tried to overcome this challenge by using virtual machines based on
Oracle's \code{VirtualBox} technology. We have, for example, developed
VMs for the Nova
project\footnote{\url{http://www.nugridstars.org/data-and-software/virtual-box-releases/copy2_of_readme}},
which is admittedly now mostly defunct. While in principle VMs allow
one to load a pre-defined software stack into a VM as well as the
VOspace access tools, in practice, this technology has not been
adopted broadly. The reasons for that were a combination of rather
time-consuming and ridged maintenance requirements and reports of
usability issues. In addition to their heavy weight nature, the VMs
also did not properly address the need for distributed teams to
collaborate on the same project space, because they limit any VM
instance to only one researcher.

From the beginning in 2007, the NuGrid collaboration had adopted
Python as the common analytic language. Around 2012/2013 the ipython
notebook technology became increasingly popular in the
collaboration. In collaboration with CANFAR, we developed as one of
two applications of the project \textit{Software-as-a-Service for Big
  Data Analytics} funded by
Canarie\footnote{\url{http://www.canarie.ca}} the
\textit{Web-Exploration for Nugrid Data Interactive}
\citep[WENDI][]{jones:14} service, which had some of the functionality
that is now offered by \code{JupyterHub}.  This project was very
successful in establishing a prototype for web-enabled analytic remote
data access with a pre-defined, stable analytic software stack and
network proximity to the data for a fast and interactive, remote data
exploration experience. For some years NuGrid served graphical
user-interfaces built on \code{ipywidgets} of the \code{SYGMA} and
\code{OMEGA} tools of the \textit{NuGrid Python Chemical Evolution
  Environment} \citep[\code{NuPyCEE},][]{2016ascl.soft10015R} as well
as the \code{NuGridSetExplorer} which allows GUI access to the NuGrid
stellar evolution and yield data sets
\citep{Pignatari:2016er,Ritter:2017wy}. The project was to a large
extent focused on improvements to the storage backend, enabling for
example VOspace to work well with indexed \code{hdf5} files, and thus
did not add authentication and access management to the service. The
usage was therefore limited to anonymous and time-limited access. The
service was deployed on virtual machines of the Compute
Canada\footnote{\url{https://www.computecanada.ca}} cloud
service\footnote{\url{https://www.computecanada.ca/research-portal/national-services/compute-canada-cloud/}}.

Another set of requirements for the \cyberhs\ facility
presented here originates from a stream of efforts in providing
enhanced data access within a collaboration, and ultimately to
external users, undertaken by the PPMstar collaboration
\citep{Woodward:2013uf,herwig:14}. The typical size of the aggregate
data volume for a single project involving three-dimensional stellar
hydrodynamics simulations is $\approx \unit{200}{TB}$, and the
collaboration would at any given time work simultaneously on two to
three projects. The simulations are performed at super-computing
centers, such as the NSF's Blue Waters computer at the NCSA in
Illinois, or high-performance computing facilities in Canada, such as
WestGrid's orcinus cluster at UBC or the cedar cluster at SFU.  Blue
Waters has relatively restrictive access requirements which make it
somewhat burdensome to add international collaborators to access the
data, especially temporary access for students. For interactive access
or processing access by collaboration members who cannot login to Blue
Waters, data has to be moved off the machine, which is not practical
since such external collaboration members do not have the required
storage facilities, and the network bandwidth is insufficient. In
addition, over decades the LCSE has developed custom and highly
optimized software tools to visualize and analyse the algorithmically
compressed data outputs of the \code{PPMstar} codes. More recently,
new tools have been developed using Python. The data exploration and
analysis ecosystem of the collaboration is heterogeneous and difficult
to maintain even for core members in view of the constantly changing
computing environments on the big clusters and the home
institutions. The challenge for this collaboration was to stabilize
and ease the use of legacy software, the very large data volumes and
the access and authentication when trying to broaden the group of
users that can have analytic and exploratory access to these large and
valuable data sets.

Based on these requirements, and through experience, we have combined
the latest technologies, including \code{Docker} and \code{Jupyter}
and designed \cyberhs, a system that allows easy deployment of a
customized virtual research environment (VRE). It offers flexible user
access management, and provides mechanisms to combine the research
area specific software applications and analytic tools with data and
processing to serve the needs of a medium-sized collaboration or user
group. At a larger scale, an architecture similar to \cyberhs\ has
been deployed, for example, by the NOAO in their \code{NOAO data lab},
and has been selected for the LSST Science
Platform\footnote{\url{https://docushare.lsst.org/docushare/dsweb/Get/LSE-319}}. \code{JupyterHub}-based
systems are also used in teaching large data science classes, such as
the UC Berkeley Foundations of Data Science
course\footnote{\url{https://data.berkeley.edu/education/foundations},
  \url{http://data8.org}}. At the University of Victoria a precursor
of the architecture described here has been used in both graduate and
undergraduate classes for the past three years. Another broad
installation of this type is the Syzygy
project\footnote{\url{https://syzygy.ca}} that allows institutional
single-sign-on to Jupyter Notebook servers across many Canadian
universities to access Compute Canada resources.

\cyberhs, although scaleable in the future, is at this point
addressing the needs of medium-sized collaborations which require an
easy to setup and maintain shared research environment.  The
\cyberhs\ software stack is available on GitHub\footnote{The multiuser
  and corehub single user is in
  \url{https://github.com/cyberlaboratories/cyberhubs}, the
  application hubs that are built on top of corehub is in the
  repository \url{https://github.com/cyberlaboratories/astrohubs}} and
the docker images are available on Docker
Hub\footnote{\url{https://hub.docker.com/u/cyberhubs}}. In
\Sect{sysarch} we describe the system architecture and implementation,
in \Sect{depl} we briefly sketch the typical steps involved in
deploying \cyberhs, and in \Sect{applications} we present the two main
deployed applications and how to add new applications. We close the
paper with some discussion of limitations and future developments in
\Sect{discussion}.

\section{System architecture and implementation}
\label{sec:sysarch}

In this section we describe the architecture and design features as
well as the implementation of \cyberhs. Here, a
\cyberh\ \textit{administrator} configures and deploys the service,
while a \cyberh\ \textit{user} is simply someone who connects to the
deployed service, and is not burdened with the details described here.
\subsection{General design features} To satisfy the requirements of the \cyberhs\
system, 
we combine the following components:
\begin{enumerate}
\item A thin user interface component  allows the users to easily interact with their virtual research environment (VRE).
\item The authentication component  ensures the right researchers are accessing the proper data, processing and software analytic tools in their VRE.
\item The docker spawner component that allows the users to spawn their selected choice of application containers and allows selection of Jupyter environment (notebook or lab). 
\item The image repository offers prebuilt container images for all components and can be expanded to host new VREs for other applications.
\item The notebook templates and interactive environments  help the users to get started in their  VREs and preparing their own analytic workflow.
\item A deployment component enables administrators, such as the data infrastructure experts in a collaboration, to deploy all of the above components with minimal effort and customization.
\end{enumerate}

The first three components are offered by \code{JupyterHub}, with some
modifications we made to the second one to allow dynamic white and
black listing for user management. The fourth is partially enabled by
the Cyberhubs Docker Hub
repository\footnote{\url{https://hub.docker.com/u/cyberhubs}}.  The
fifth and sixth components, and the customization images are offered
by \cyberhs.

\cyberhs\ provides a complete, packaged system that can easily be
installed and customized and allows the \cyberhs\ administrator to
quickly deploy VREs. There is no reason why a particular
\cyberhs\ instance could not be near-permanent, with little in
maintenance needed because the system is based on docker
images. However, the \cyberhs\ system is designed for easy deployment
with all essential configuration options, like attaching local or
remote storage volumes, provided through external configuration files
that then will be absorbed by the pre-built docker images.
 
\begin{figure}[]
\centering
\includegraphics[scale=0.60]{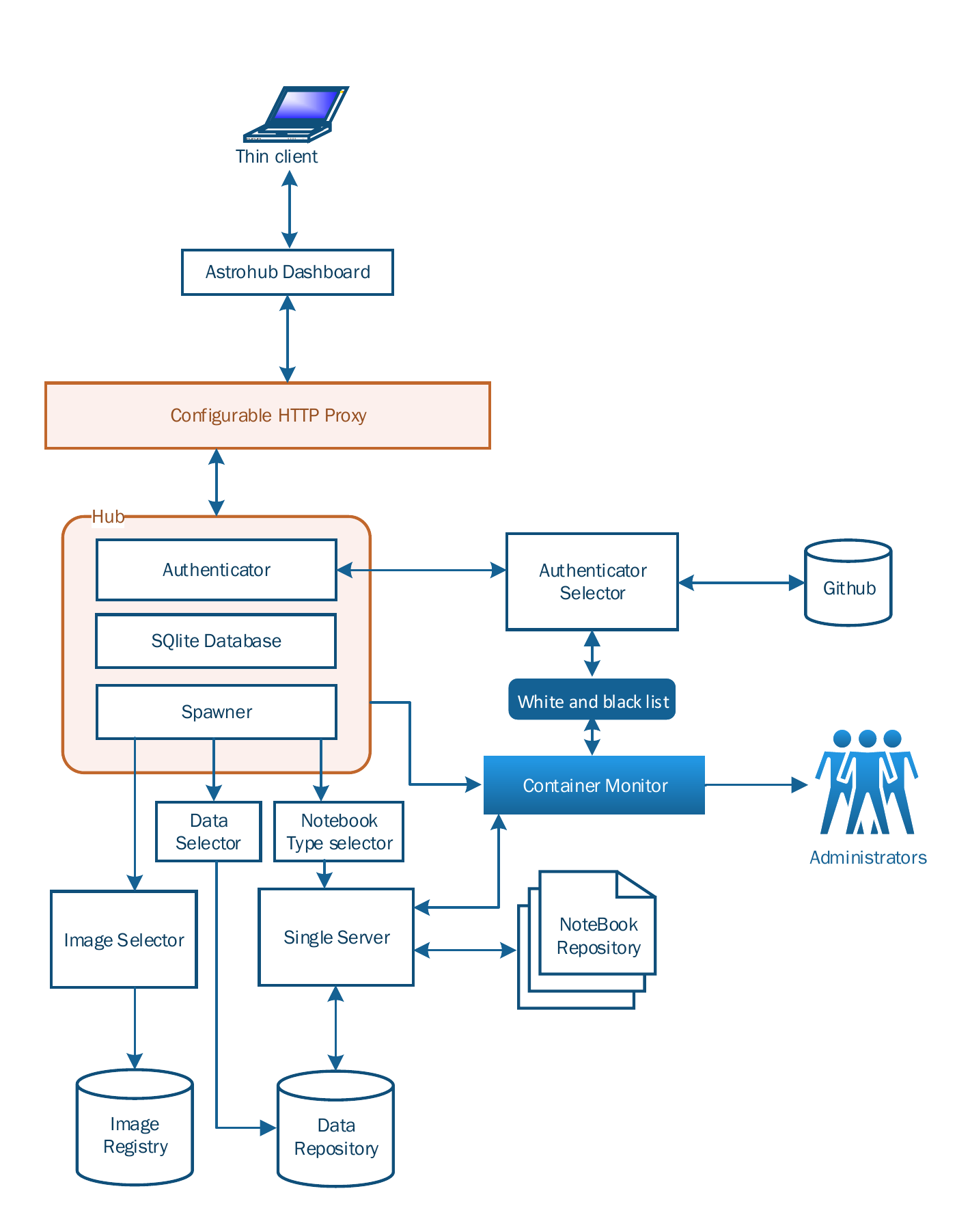}
\caption{\label{fig:astro_arch}
\cyberhs\ general system architecture.}
\end{figure}
As opposed to other packaging solutions that rely on tools such as
\code{ansible}, puppet and others, \cyberhs\ requires only setting a
minimal number of environment variables to deploy any of the pre-built
application hubs which could possibly enable many astronomy use
cases. Extending an application hub, or building a new one on top of
\code{corehub} (part of \cyberhs\ repository) is straight forward and
well documented. An example if the application \code{SuperAstroHub}
(available in the {Cyberhubs Docker repository}) available on the
WENDI server (see \Sect{wendi}) that combines all of our presently
available cyberhub applications.

The general system architecture for the \cyberhs\ is shown in
\Fig{astro_arch}.  \code{JupyterHub} is a multi-user environment that allows
the users to authenticate and launch their own notebook and terminal
server, while sharing access to certain storage areas.  By doing so,
the same collaboration infrastructure can be used by multiple users
and therefore provide a platform for sharing resources, analytic
tools, as well as research content and outcomes.  The main components
of \code{JupyterHub} are:
\begin{itemize}
  \item A configurable HTTP proxy that allows the users to interact
    with the system and directs their requests to the appropriate
    service.
  \item A hub that handles users and their notebooks. In more detail,
    the hub offers the following services:
   \begin{itemize}   
   \item An authentication service that supports many authentication
     backends (including PAM, LDAP, OAuth, etc.). Currently,
     \cyberhs\ uses an extended Github authenticator that allows users
     with GitHub account to log into \cyberhs.
     \item A spawner service that allows the user to select the
       singleuser application hub and the interface type.
   \item An SQLite database which keeps track of the users and the
     state of the hub.
 \end{itemize}  
\end{itemize}

\subsection{Spawner and authentication extensions}
Our extended spawner is integrated in the \code{JupyterHub} configuration
file. It provides at this point two selection options. First the user
selects between the default Jupyter Notebook option and the
experimental option JupyterLab.  Both JupyterLab and Notebooks offer
Python, bash and other language notebook options as well as terminal
access to the singleuser hub container. JupyterLab is a significantly
enhanced Jupyter interface that overcomes the restrictions imposed by
single linear notebooks or terminals, and allows one to combine
multiple sessions in parallel in one web browser window. For the
second option, the user can select the application hub image if
multiple options are configured to be offered by the
\cyberhs\ administrator. In the future a simple extension could allow
one to choose between a variety of data access options in this spawner
menu.

A major challenge in a shared environment is access and authentication
administration.  For most collaborations, national or institutional
authentication models are not practical. We have adopted the
third-party OAuth authentication method available for \code{JupyterHub} and
allow authentication of users with their GitHub accounts. Other
third-party OAuth applications, such as Google, could be used as well.

To dynamically control the GitHub users who can access the system, our
authentication extension provides a simple whitelist and blacklist
mechanism that can be updated in the running, fully deployed
\cyberhs\ without service interruption. The authenticator relies on a
whitelist and/or blacklist file to dynamically grant or deny access to
GitHub users. When a whitelist is supplied, only the users in that
list are allowed to login. When the blacklist is present, the users in
the blacklist are blocked from accessing the system and will get a
\code{403: forbidden} page even if they are in the whitelist.

This is a rather simple yet powerful access model that allows, in
combination with the easy configuration of the storage additions to
the \cyberhs, a flexible access control to data and processing that can
serve in a flexible way the access and sharing requirements of
medium-sized collaborations. It can be easily combined with temporary
unrestricted access to any user.
 
Unlike some \code{JupyterHub} installations that propagate the host system
identity to the hub user, or systems that create inside the
singleuser application container an identity according to the
login identity, we are adopting a simpler approach with the goal of
enabling the most transparent and seamless sharing and
collaboration. In \cyberhs\ all users have in their own application hub
container the identity \textit{user}. Typically, a read-write data
volume of considerable size is attached, and all users appear as users
on that read-write volume with the same identity.

Another major component in a shared environment is resource
allocation.  By resources we are referring to cpu, memory, swap space
and disk storage.  Currently, we are not enforcing any resource limits
and we are not offering any scheduling capabilities.  We are
considering to add the abilities to specify resource limits for every
container launched and to alert users if no more resources are
available.  Since these abilities do not ensure a scalable system, our
roadmap includes a plan to use Docker Swarm and/or Kubernetes for
scaling the resources and scheduling the containers.

\subsection{Features and capabilties}
Both Jupyter Notebook and Jupyterlab offer web-based notebook user
interfaces for more than 50 programming languages, and we included by
default python 2 and 3 and bash. But other popular languages, such as
R or Fortran are easily added. Both bash notebooks as well as
terminals provide shell access to the singleuser application docker
container (that we call here hub). Any simulation or processing
software that can be executed on the Linux command line, such as the
\mesa\ stellar evolution code
\citep{Paxton2011,Paxton:2013km,Paxton:2015iy} or the NuGrid
simulations codes \citep{Pignatari:2016er}, can be run by each user in
their identical instance on the full hardware available on the
host. Other examples include legacy analysis and processing tools that
require a special software stack. If such software is once expressed
in the singleuser application hub docker image it can be easily shared
with anyone accessing the \cyberhs.

\cyberhs\ are typically configured with an openly shared, trusted user
space that is equally available to all users that have access to a
particular \cyberh. This user space is mounted on a separate, persistent
volume. A trusted collaboration, would establish some common sense
rules on how to access this shared space, in which all participants
have seamless access to any project or individual directories.  We
have found that this arrangement allows for very effective cooperation
between team members with very different skill sets, including
students.

A typically small amount of private non-persistent storage is
available inside the user's singleuser hub instance, which will
disappear when the user container is restarted. In order to create
some level of persistence beyond the shared user space, users of
\cyberhs\ rely heavily on external, remote repositories, such as git
repositories, for storing and sharing non-data resources, such as
software, tools, workflows, documentation, and paper writing
manuscripts.

In addition to analytic access to data the \cyberhs\ provide
documentation and report writing through inline Markdown cells, a
latex typesetting environment where pdf files are seamlessly viewed in
the browser for paper manuscript writing and editing, as well as slide
presentation extensions to Jupyter which allow one to create
presentations with live plots and animations. In addition, graphical
user-interface applications can be built, and examples are provided in
\code{wendihub} (\Sect{wendi}).

In terms of the maintenance of the multiuser environment, where many
containers are running, the \cyberhs\ administrator should monitor the
state of the \cyberh, including the available resources.  Astray and
blacklisted user containers should be removed.

\begin{figure*}[th]
\centering
\includegraphics[width=1.8\columnwidth]{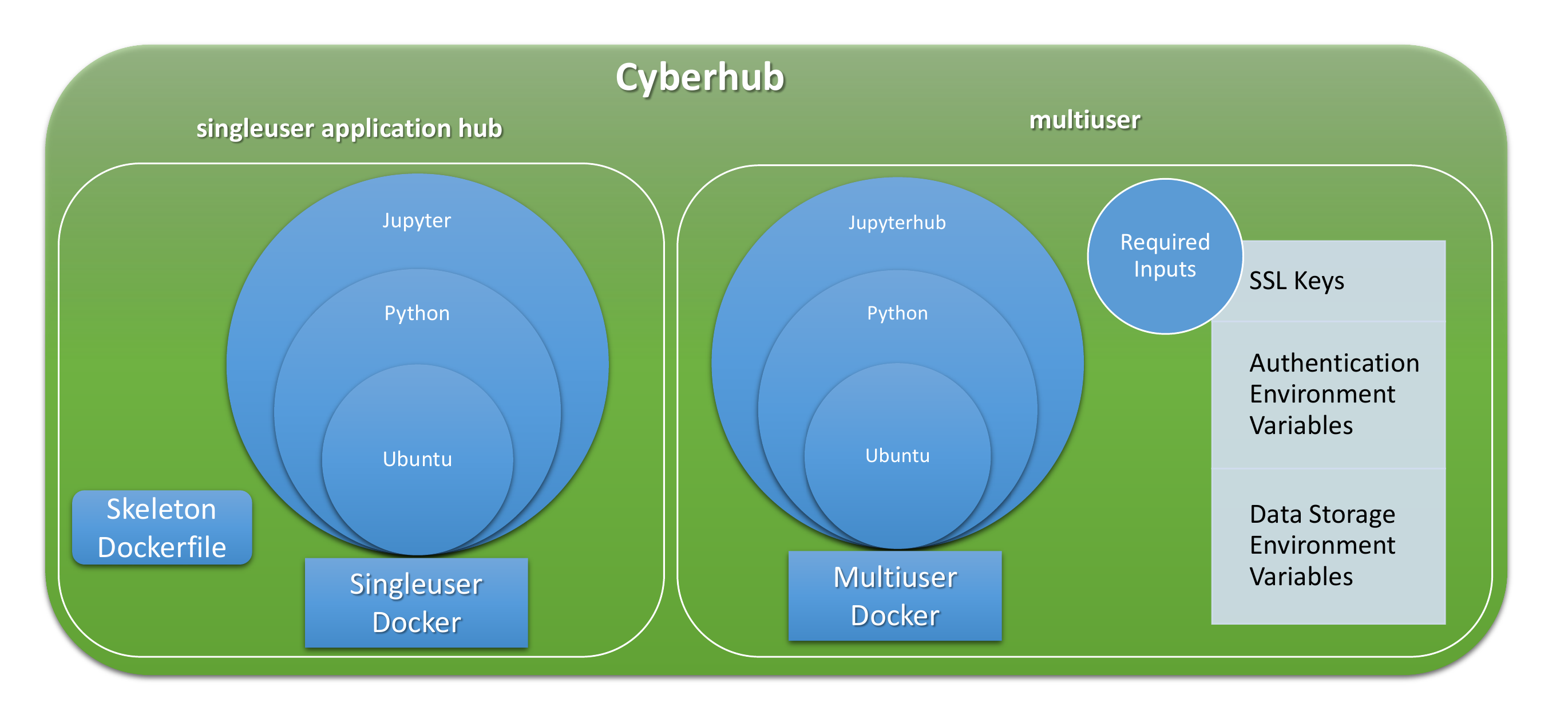}
\caption{\label{fig:corehub_comp} Main elements of the core
  \cyberhs\ system. The \code{multiuser} takes care of receiving the
  initial service request from a user, handles user authentication and
  and data storage attachment. It launches a singleuser application
  hub from the appropriate image or reconnects a returning user to
  that user's existing application container. The singleuser component
  contains the application-specific analytic software and is the
  processing home for the user. Each user has a separate container
  instance of the application image.}
\end{figure*}
A typical \cyberh\ includes a repository of examples and template
notebooks to help users getting started in exploring the data
resource. These notebooks can be copied into the image or provided via
mounted volumes. Each \cyberh\ in our \cyberhs\ family has strict
version-specific requirements files for python and Linux packages,
that ensure that the same versions of each component of the entire
software stack are always used, until an update is made. In that case,
past docker image versions will be still available as tagged images on
the docker hub repository. Each user has therefore a completely
controlled and reproducible environment. It is then straight-forward
to create a stable, shareable and reproducible science
workflow. Simulation software and analysis packages are shared via
repository platforms such as GitHub, GitLab or BitBucket, and include
information on exactly which \cyberh\, including the version, it is to
be deployed.
 
At the core of the \cyberhs\footnote{These are available in the GitHub
  repository \url{https://github.com/cyberlaboratories/cyberhubs}.}
design is the multiuser image and a basic \code{corehub} singleuser
application. The latter is a skeleton and has no application software
installed.  The main elements of these core elements of \cyberhs\ are
shown in \Fig{corehub_comp}. They are:
\begin{enumerate}
\item multiuser image: It is composed of our customized \code{JupyterHub}
  image, available via dockerhub repository, or via build package that
  includes Dockerfile and all other necessary scripts and
  docker-related files to customize, compose and then launch the
  multiuser {JupyterHub} service.
\item singleuser \code{corehub} image: The most basic, bare singleuser
  docker image, also available via dockerhub repository or build
  package.
\end{enumerate}  
\code{corehub} is the starting point on top of which all of the other
application hubs are built (\Sect{depl}), as shown in
\Fig{corehubs}. Obviously, not only astronomy \cyberhs\ can be built on
top of \code{corehub}, but applications from other disciplines or use
case are possible. The astronomy application hubs are the
\code{astrohubs}\footnote{They are available from the GitHub
  repository \url{https://github.com/cyberlaboratories/astrohubs}. As
  explained in the documentation provided with these repositories the
  docker images are staged at
  \url{https://hub.docker.com/u/cyberhubs}.}.

\subsection{Storage staging}
Docker containers typically do not have much storage, and a few
options of storage staging are typically deployed:
\begin{itemize}
\item {Read-only data volume}: Most \cyberhs\ are about providing access
  to a particular data universe. This is immutable data that is staged
  on read-only data volumes. It is mounted on the singleuser container
  to allow the users to read the data in their notebooks and
  processing as needed.
\item {Persistent data volume}: This volume is also mounted and all
  users have the ability to write to and read from it. The volume
  lives on the host or externally attached storage, and is protected
  against singleuser container shutdowns.
\item {Local ephemeral storage}: This is the local storage allocated
  to each container when created. It is available to the users only
  when the container is running and gets purged once the container is
  removed. This holds a copy of example notebooks added to the image
  to be available to all users. This it the home directory of each
  user. Since this area is inaccessible to other users of the
  \cyberh\ it is the right place to store, for example, a
  \code{.gitconfig} file or other configuration files as well as a
  \code{.ssh} directory.
\item {Individual remote data storage}: Users can use sshfs, mountvos,
  google-drive-ocamlfuse, and other fuse tools to mount their remote
  data storage. This, however, requires elevated privileges for the
  containers and is not yet supported.
\end{itemize}
Currently, read-only and persistent data volumes are specified via
environment variables, and we do not support individual remote data
storage yet. However, the \cyberhs\ administrator can now configure both
read-only and persistent data volumes as remote data volumes, an
option that already would serve the needs of many collaborations.  We
are considering to add a data selector that allows users to select
volumes to mount that they have privilege for.

\section{System Deployments}
\lSect{depl} The goal of \cyberhs\ is to make deployment as easy as
possible, and there are two deployment options. \cyberhs\ is based on
dockers whose images are created and stored in a repository either
locally, or on the \code{Docker Hub
repository}. A Docker container
is an instance of a Docker image. A \cyberh\ deployment always consists
of two Docker containers (\Fig{corehub_comp}). One is an instance of
the \code{cyberhubs/multiuser} image, the other is one container per
user of one application hub image, such as \code{cyberhubs/corehub}. A
particular configuration may choose to offer users more than one
application image in the spawner menu dialogue after login, such as
those offered in the \cyberhs\ family.

For ease of use, the first deployment option is recommended. It
involves launching containers as instances from the images available
in the \cyberhs\ organization at
\url{https://hub.docker.com/u/cyberhubs} which allows deployment of a
\cyberh\ without building any docker images. The only requirement is
one must prepare the host machine and the \cyberh\ configuration file
and launch.

The second option is for the administrator to build one or several of
the required Docker images using the Dockerfile and configuration
files provided. This involves building the \code{singleuser}
application hub on top of one of the provided application hub
images. This option allows the administrator to add features and
software otherwise not available by default, and to provide
specialized configurations.

For specializations that require fundamentally different data access
or different user authentication models, it may be necessary to
rebuild the \code{multiuser} image.
\begin{figure}[ht!]
\centering
\includegraphics[width=\columnwidth]{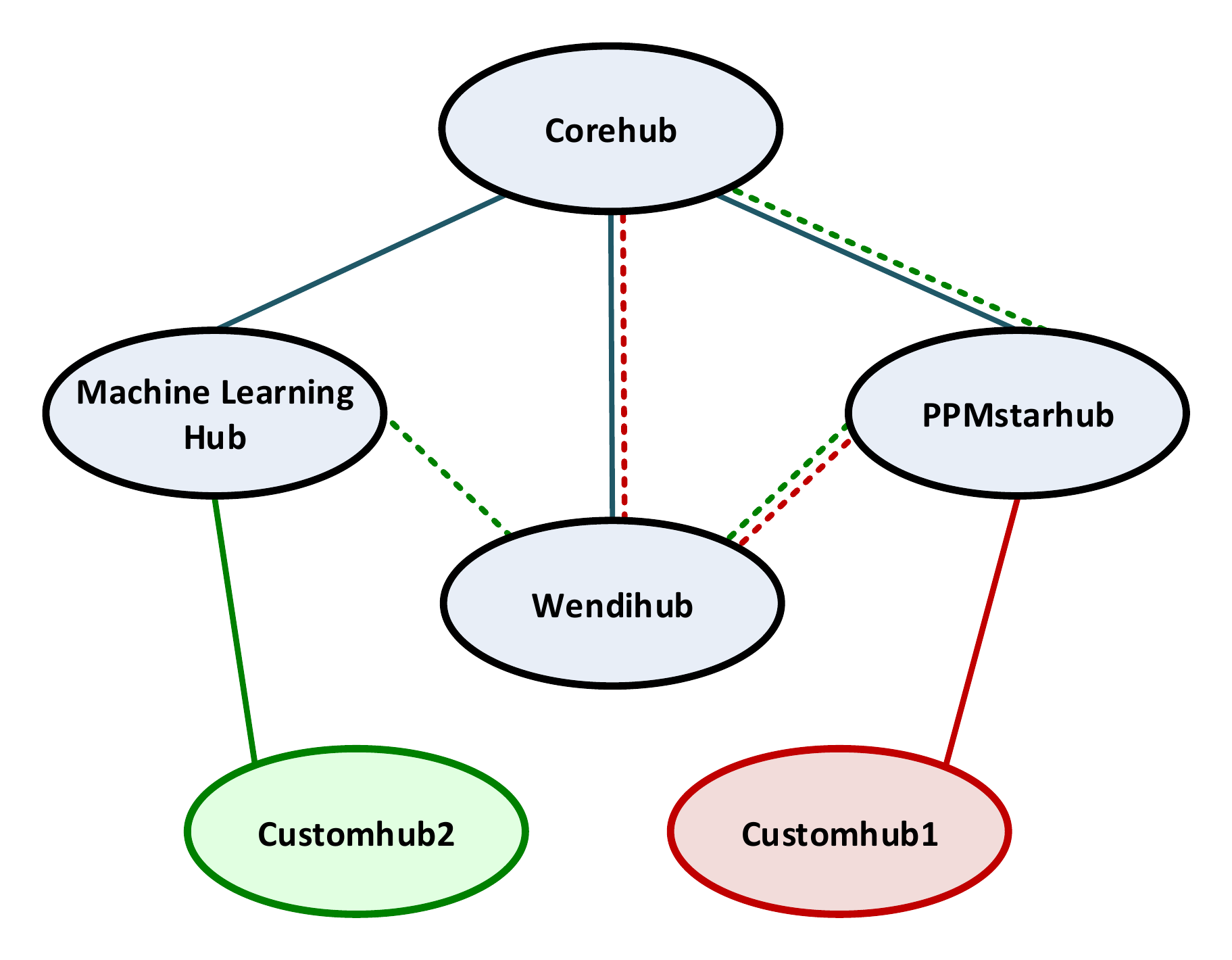}
\caption{\label{fig:corehubs} \cyberhs\ administrators can build
  application images by chaining existing application images and
  adding customization through dockerfile-based builds.  By starting
  from the pre-built \code{PPMstarhub} that is itself based on
  \code{Codehub} one can build in sequential steps \code{Wendihub} and
  \code{Machine Learning Hub} each on top of the previous. At this
  point the capabilities and analytic tools of three application hubs
  are combined, and can be the bases for another addition of tools and
  data stores to create \code{Customhub2}. Similarly, for example,
  \code{Customhub1} can be built as a combination of \code{Corehub},
  \code{Wendihub} and \code{PPMstarhub}.}
\end{figure}
In any case the administrator follows the step-by-step documentation
in the
\cyberhs\ repository\footnote{\url{https://github.com/cyberlaboratories/cyberhubs}}
on GitHub. A deployment would involve the following steps:
\begin{itemize}
\item Preparing the host machine:
\begin{itemize}
  \item Select a host machine, such as a Linux workstation. In our
    case \cyberhs\ are deployed in a cloud environment, such as the
    Compute Canada Cloud, and require launching a suitable virtual
    machine, attaching external storage volumes, and assigning IPs. We
    use a CentOS7 image, but other Linux variants should work as well.
  \item The host machine only needs a few additional packages, most
    important of which is the \code{docker-ce} package.  A small
    amount of docker configuration is followed by launching the docker
    service on the host machine.
  \item Any external data volumes to be made available need to be
    mounted. \code{sshfs} mounted volumes work well.
\end{itemize}
\item Configuring the \cyberhs. This invariably starts with pulling the
  \cyberhs\ Github
  repo\footnote{\url{https://github.com/cyberlaboratories/cyberhubs}}. The
  main steps that always must be done are:
\begin{itemize}
  \item Register an OAuth authentication application with a GitHub
    account and enter the callback address, authentication ID and
    secret into the single configuration file
    \code{jupyter-config-script.sh}.
  \item Specify the admin user IDs as well as white-listed users. If
    white-listed users are specified, either in the configuration file
    (static white listing) or in the \code{access/wlist} file (dynamic
    white listing), then only white listed users are
    allowed. Otherwise everybody with a github account can have
    access. Dynamic black listing is also possible once the service
    runs.
  \item Specify or modify data storage mapping from the host to the
    \cyberh\ as the user sees it, for both read-write and read-only
    storage.
  \item Specify the application hub name. Either a single application
    hub is offered or the spawner menu can offer a number of different
    application hubs.
  \item Finally, create SSL key/certificates as described in
    \code{multiuser/SSL/README}. While a commercial certificate can
    certainly be used, our default installation includes
    \code{letsencrypt} which provides free three-month
    certificates. Such certificates can be created and updated using
    the docker \code{blacklabelops/letsencrypt}. However, the present
    instructions recommend the use of
    \code{certbot-auto}\footnote{\url{https://dl.eff.org/certbot-auto}}
    which works well for our reference host system CentOS.
  \item Source the config file and launch the \cyberh\ according to the
    instructions in \code{multiuser/README}.
\end{itemize}
\item The above assumes a deployment from the pre-built docker
  images. This is the recommended mode. By default, the
  \code{multiuser} image will be automatically pulled from the
  repository during the execution of the \code{docker-compose up}
  command. However, the singleuser application hub will have to
  be pulled manually using the \code{docker pull} command. In addition
  to the basic \code{corehub} application there are several pre-built
  applications available in the docker hub
  repository\footnote{\url{https://hub.docker.com/u/cyberhubs}}, such
  as WENDI for NuGrid data analysis, \code{mesahub} for \code{MESA}
  stellar evolution, for machine learning hub \code{mlhub} to be used,
  for example, by StarNet \citep{Fabbro:2017vx}, \code{PPMstarhub} for
  \code{PPMstar} stellar hydro data analysis.
\item To add more application functionality, packages, tools, etc.\ it
  is necessary to rebuild the singleuser application hub. The Docker
  and configuration files of all of the existing application hubs
  available on Docker Hub are in the \code{astrohubs} GitHub
  repository\footnote{\url{https://github.com/cyberlaboratories/astrohubs}}. All
  of these application hubs start with the \code{cyberhubs/corehub}
  image, as indicated in the first line of the Dockerfile: \code{FROM
    cyberhubs/corehub}.  Building or extending an application hub
  could start from one of the existing hubs, or from
  \code{corehub}. Within the \cyberhs\ family, all application hubs can
  be combined with all others. This is shown schematically in
  \Fig{corehubs}. In principle a super-application hub can be built by
  daisy-chaining all other application hubs together, collecting and
  adding in the process the capabilities from each participating
  application hub. If the community builds application hubs that are
  consistent with the \cyberhs\ model, they could be added to the
  \cyberhs\ docker hub repository.
\item The \code{multiuser} image and all application hub images can
  also be built from scratch using the dockerfiles and configuration
  files provided in the GitHub repositories, starting with
  \code{Ubuntu} images. This allows full customization of all
  components, or improvements of the \cyberhs\ facility. The complete
  rebuild of the docker images would also be required when an update
  of the software stack is needed.
\end{itemize}

\section{Applications}
\lSect{applications} In this section we describe the application hubs
we have built and deployed.  Due to our research areas, these enable
simulation-based data exploration.  Applications for
observationally-oriented data exploration and processing tasks are
equally possible and supported.

\subsection{NuGrid}
\lSect{nugrid_application}
As described in \Sect{intro}, the challenges of the NuGrid
collaboration have contributed a significant portion to the
requirements of \cyberhs.  NuGrid develops and maintains a number of
simulation codes (\code{NuPPN}) and utilities that allow users to
perform nucleosynthesis production simulations and nuclear physics
sensitivity studies. These codes, the \mesa\ code
\citep{Paxton2011,Paxton:2013km,Paxton:2015iy} and the GENEC stellar
evolution code \citep{Eggenberger:2008} have been used to create the
NuGrid data sets \citep{Pignatari:2016er,Ritter:2017wy}.  The NuGrid
data consists of stellar evolution tracks for twelve initial masses
and five metallicities. Each track is made up of between 30,000 and
100,000 time steps. For each time step profile information including
at least density, temperature, radius, mass coordinate, mixing
coefficient, and a small number of isotope abundances is saved. Each
profile has between 1,000 and 5,000 radial zones, and for all profiles
all zones are written out. The data set also contains post-processed
data which reports profiles for about 1000 isotopes every 20 time
steps.

\begin{figure}[hbt]
  \includegraphics[width=\columnwidth]{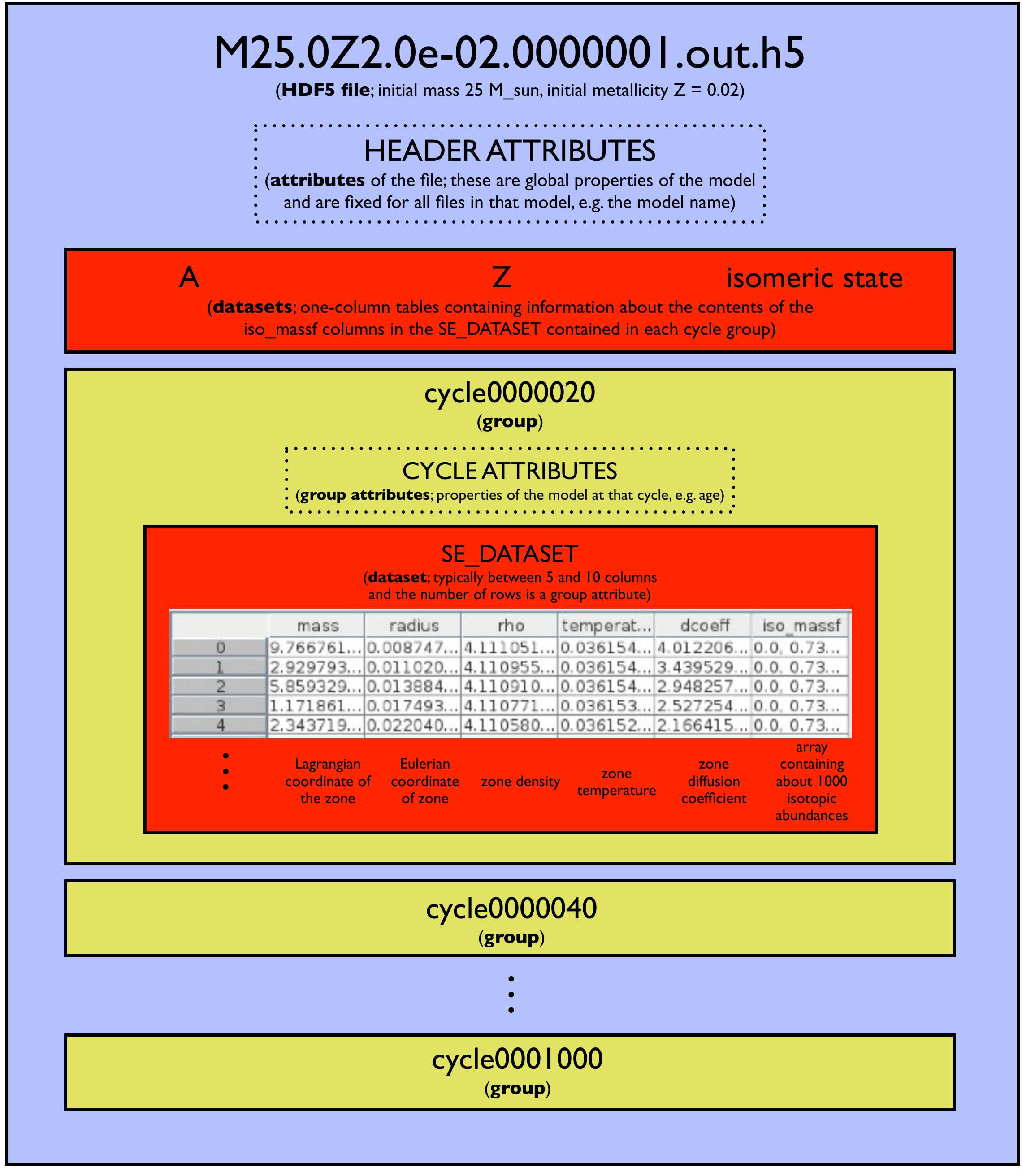}
  \hspace*{0.2\columnwidth}\includegraphics[width=0.6\columnwidth]{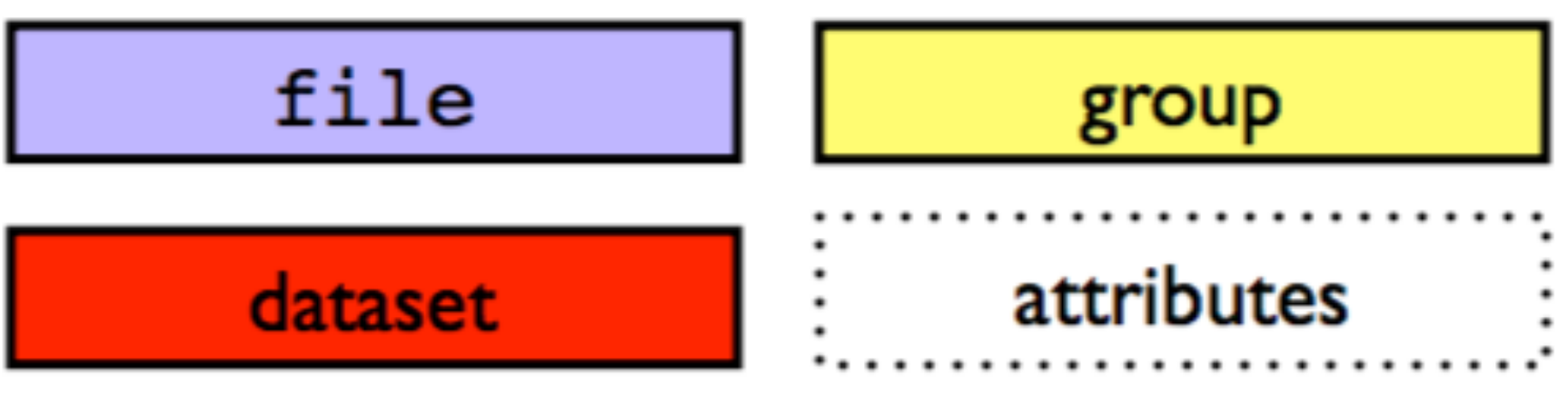}
  \caption{\lFig{seformat} Schematic of SE data format for
    one-dimensional time-dependent stellar evolution and explosion
    data.}
\end{figure}
The time-dependent nature of stellar evolution simulation output
suggests for this data a particular structure.  For each saved time
step, or cycle, a number of scalar quantities have to be saved, as
well as a number of profile vectors. A number of such cycles are
combined into one data file, or \textit{packet}. The scalar quantities
are the \textit{cycle attributes}, the profile vectors are the
\textit{data columns}, and each \textit{packet} has a number of
\textit{header attributes} which are repeated in each file and provide
global information for the run, such as initial conditions, code
version used, units of quantities in the data columns or cycle
attributes, etc. This is the SE (Stellar evolution) data format, shown
schematically in \Fig{seformat} and, within NuGrid, is currently
implemented using the \code{hdf5} data format.

\code{SE} output from \mesa\ is written with NuGrid's
\code{mesa\_h5} and
then used for the post-processing simulations using NuGrid's
\code{NuPPN} codes, which in turn write output again in the \code{SE}
format. Libraries and modules to write \code{SE} from \code{Fortran},
\code{C} or \code{Python} are
available\footnote{\url{https://github.com/NuGrid/NuSE}}. \code{SE}
data output can be explored via data access, standard plots,
visualisations, and standard analysis procedures, using the
\code{NuGridPy}\footnote{\url{https://nugrid.github.io/NuGridPy}}
Python package.

We want to accommodate three types of user:
\begin{itemize}
  \item Internal users, including for example students, who are less
    experienced in the analysis of the NuGrid simulation data.
  \item Users external to the collaboration to whom we would like to
    provide the option to explore the NuGrid data set to find answers
    to their specific research questions.
  \item Expert users who want to carry out NuGrid and/or
    \mesa\ simulations and analyse the simulation output conveniently
    in the same location, and possibly share run directories and
    workflows. These users can share a common development platform
    that is exactly identical to each participant who remotely
    accesses the platform and all of its content.
\end{itemize}
The first two types of users are served by WENDI, while the third
requires additional compilers and libraries which are delivered in
\code{mesahub}.

\subsubsection{WENDI}
\lSect{wendi} Web-Exploration for NuGrid Data Interactive
\code{WENDI} provides
\begin{itemize}
	\item python and bash notebooks, terminal and text editor;
	\item example notebooks; and
	\item self-guided graphical user interface (GUI) notebooks
          (\textit{widgetized} notebooks).
\end{itemize}

The WENDI widget notebook \textit{NuGridStarExplorer} provides GUI access to
plotting and exploring the NuGrid data sets, specifically the library
of stellar evolution and detailed nucleosynthesis simulations.
As an example, the evolution of a low-metallicity,
  intermediate mass star is shown during the Asymptotic Giant Branch
  evolution. The Kippenhahn diagram shows the Lagrangian coordinates
  of recurring He-shell flashes, each of which drives a \pdcz. In the
  \pdcz\ the $\isotope{}{22}{Ne}(\alpha,\neutron)\isotope{}{16}{O}$
  reaction creates neutrons with neutron densities reaching $N_n
  \approx 10^{12}\mathrm{cm^{-3}}$. At these neutron densities
  \spr\ branchings, such as at \isotope{}{95}{Zr}, are activated and
  the neutron-heavy \isotope{}{96}{Zr} is
  produced. \Fig{WENDI_iso_abu} shows the isotopic abundance
  distribution of s-process elements. Note, how the mass fraction of
  \isotope{}{96}{Zr} exceeds that of \isotope{}{94}{Zr} in the shown
  model at the end of the \pdcz. In the \isotope{}{13}{C} pocket where
  the bulk of the s-process exposure takes place
  $\isotope{}{94}{Zr}/\isotope{}{96}{Zr} \approx 600$ because the
  neutron denisty is low and the branching is closed. In the solar
  system $\isotope{}{94}{Zr}/\isotope{}{96}{Zr} = 6.2$. This
  demonstrates the different neutron density regimes of the
  \sprn. Although this example is for low metal content, conditions at
  solar-like Z are similar. The solar system abundance distribution
  originates from a mix of low- and intermediate mass stars, involving
  both the \isotope{}{13}{C} and the \isotope{}{22}{Ne} neutron
  source, which each produce isotopic ratios in different
  proportions \citep{gallino:98,herwig:04c,Herwig:2013fc}.
\begin{figure}[hbt]
  \begin{center}
    \includegraphics[width=0.85\columnwidth]{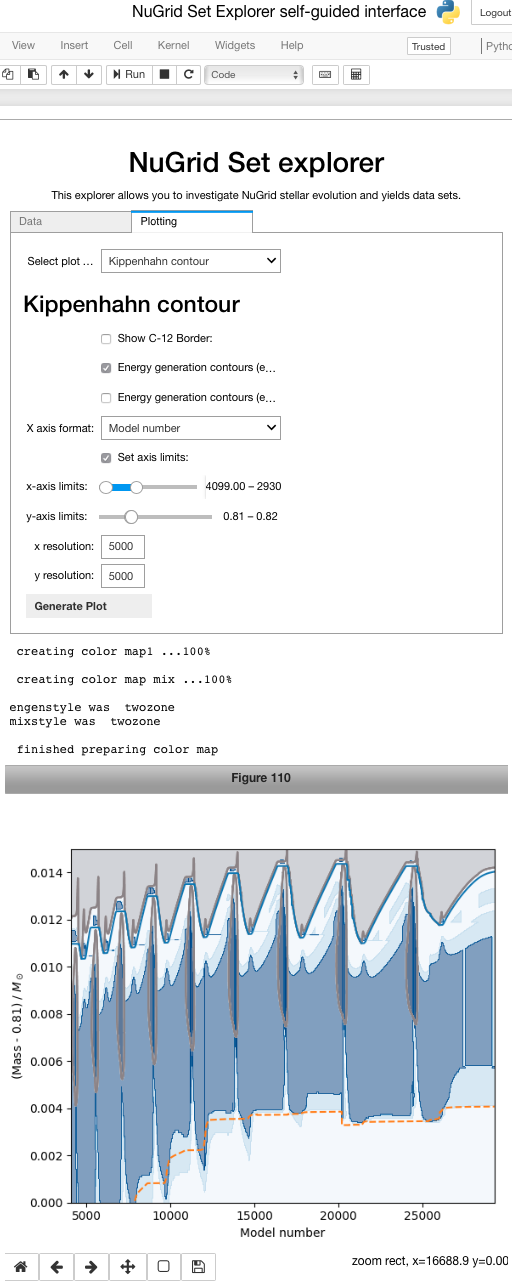}
    \caption{Self-guided exploration of the NuGrid stellar evolution
      and yield data base via the graphical user interface
      \textit{NuGridStarExplorer} in WENDI. Kippenhahn diagram of a
      $\Mzams=3\Msun$ stellar model with $Z=0.001$, zoomed in around
      the core-envelope interface where the He- (dashed, orange line)
      and H-burning (solid, blue line) shells are located. Grey and
      blue areas mark convectively unstable regions and regions of
      energy generation. The isotopic abundance distribution in the
      thermal pulse ending around model 16880 is shown in
      \Fig{WENDI_iso_abu}.  } \lFig{WENDI_kip_contour}
  \end{center}
\end{figure}
\begin{figure}[hbt]
  \includegraphics[width=\columnwidth]{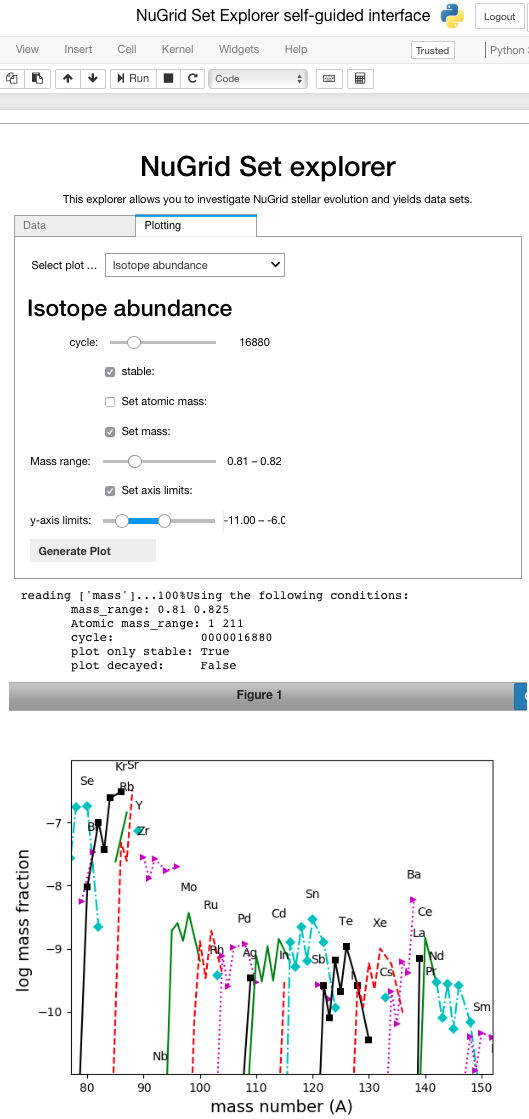}
  \caption{Mass-averaged isotopic abundance distribution of the
    \pdcz\ for model 16880 of the $\Mzams=3\Msun$ stellar model with
    $Z=0.001$ shown in \Fig{WENDI_kip_contour}. The abundance
    distribution shows the stable isotopes for the first- and
    second-peak \spr\ elements (see text for more details).}
  \lFig{WENDI_iso_abu}
\end{figure}

While the widget notebooks provide easy and powerful access to the
data, customized analysis may require the added control of programming
access to the platform. In order to make it easy for users to get
started analyzing the NuGrid data we keep adding to a
collection\footnote{\url{https://github.com/NuGrid/wendi-examples}} of
short example analysis tasks, such as \textit{abundance profiles at
  collapse}, or \textit{C13-pocket analysis}. All data and software
dependencies of these examples are satisfied on
\code{wendihub}. Although users can easily clone their own copy of any
external repository, the \code{wendi-examples} are preloaded in WENDI,
and are a convenient starting point for further analysis. Users who
create an interesting new example are encouraged to fork the example
repository on the terminal command line, add their new example and
make a pull request to the original repository.  Although executing
notebooks is just a matter of clicking the play button, any further
interaction for this type of notebook-based analysis requires basic
knowledge in Python.

Two additional widget
notebooks\footnote{\url{https://github.com/NuGrid/WENDI}} are
presently available in WENDI. The \code{OMEGA} self-guided interface
provides example applications of the NuGrid Python Chemical Evolution
Environment \citep[NuPyCEE,][]{2016ascl.soft10015R} code One-zone
Model for the Evolution of Galaxies (OMEGA), such as models for dwarf
galaxies Fornax, Carina and Sculptor.  For OMEGA WENDI allows
arbitrary, complex programming of analysis through ipython notebooks
as well. For example, the top panel of
\Fig{figure_science_case_OMEGA_SYGMA} shows that NuGrid massive star
yields overproduce some iron-peak elements like Cr and Ni, but produce
a consistent amount of other elements like Ti, V, Cu, and Ga.  Further
analysis of the yield source as shown in the IMF-weighted yields
(bottom panel) reveals that the overestimation of Cr in the galaxy
evolution models originates from the $20 \Msun$ model at $Z=0.01$.
This analysis tool allows to identify sources of discrepancies between
numerical predictions and observations. In this case, further analysis
of the underlying stellar evolution model using similar approaches as
those demonstrated below for intermediate-mass stars, that will be
presented elsewhere in more detail, connects this overproduction of
some iron-group elements to the convective merger of an O-Si shell in
the stellar evolution model that may not be realistic. This analysis
can be performed by anyone on WENDI, and is available there in a
notebook as part of the pre-loaded WENDI
examples\footnote{\url{https://github.com/NuGrid/wendi-examples/blob/master/Stellar\ evolution\ and\ nucleosynthesis\ data/Examples/Solar_abundance_distribution.ipynb}}.
\begin{figure}[hbt]
  \includegraphics[width=\columnwidth]{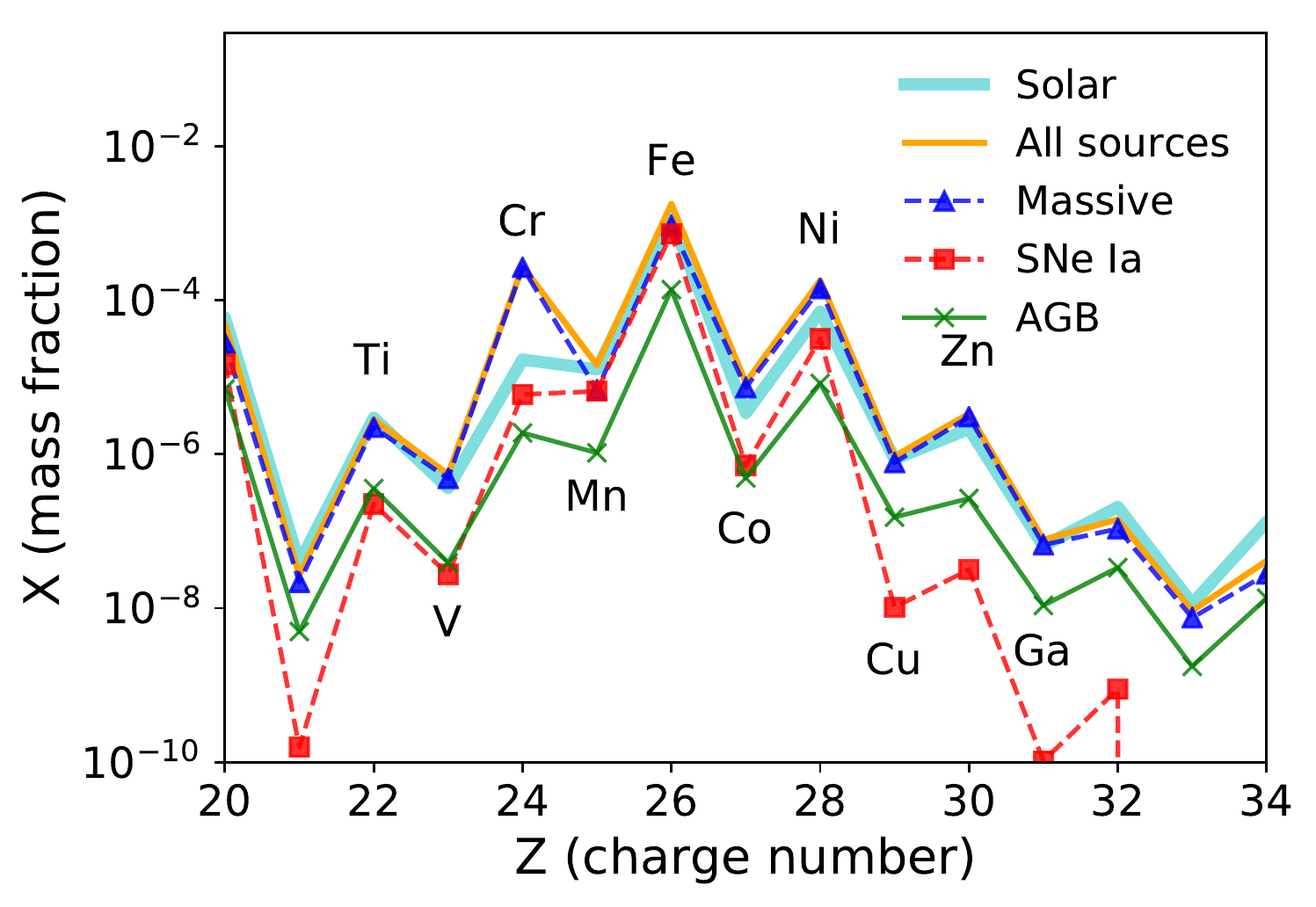}
  \includegraphics[width=\columnwidth]{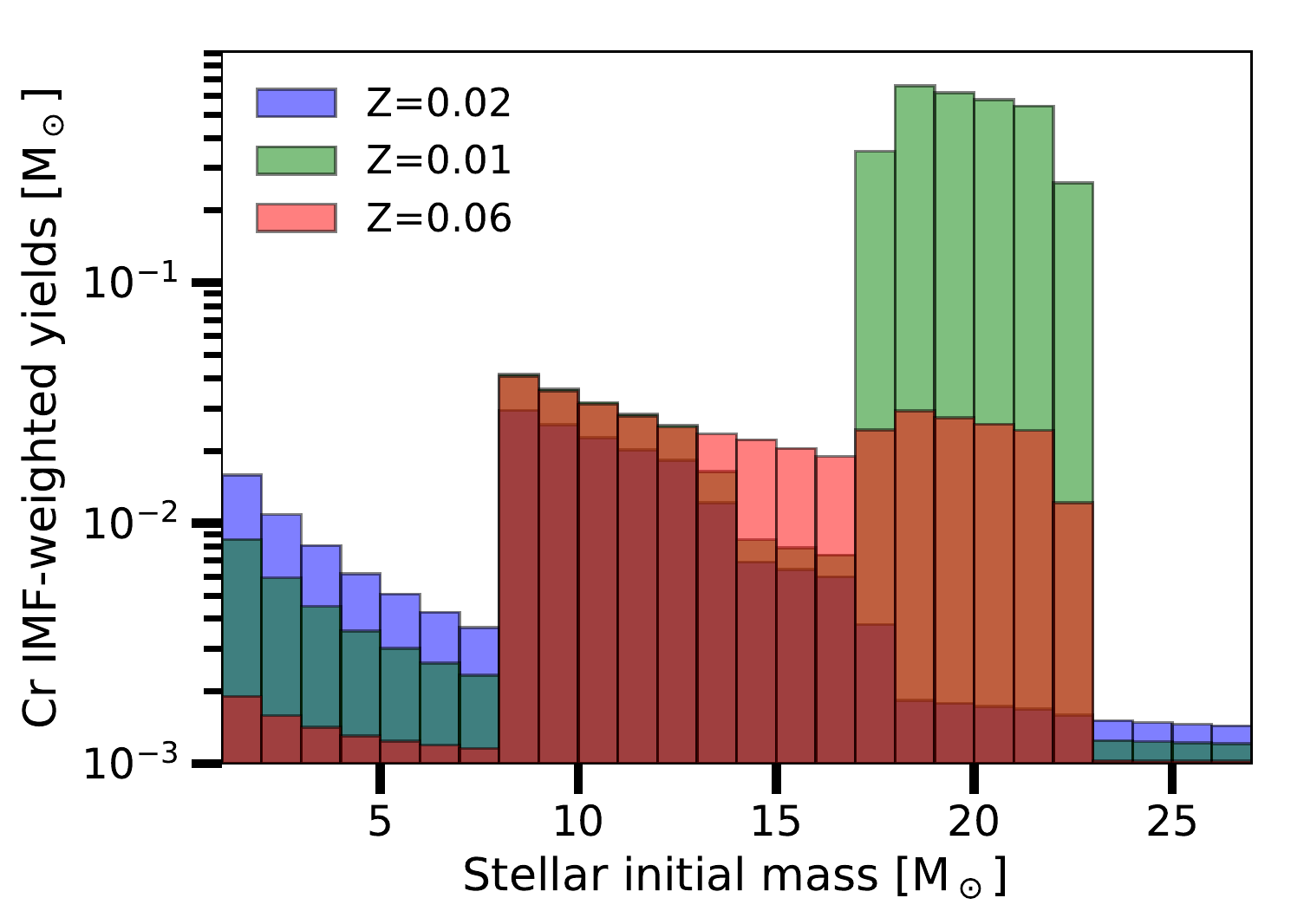}
  \caption{{\bf Top:} Elemental abundance distribution of the Galactic
    gas, when the Sun formed, predicted by the chemical evolution code
    OMEGA using NuGrid yields for low-mass and massive stars and
    \cite{thielemann:86} yields for Type-Ia supernovae (SNe Ia). The
    solar distribution is taken from \citet{Lodders:2009er}. The blue
    (dashed with triangles), red (dashed with squares), and green
    (solid with crosses) lines represent the individual contribution
    of massive stars (winds and core-collapse supernovae), SNe Ia, and
    low-mass stars, respectively. The orange solid line shows the
    combined contribution of all sources. {\bf Bottom:} Cr yields
    weighted by the initial mass function as a function of stellar
    mass for different metallicities (different colors) in a $10000
    \Msun$ simple stellar population, using NuGrid yields and the
    SYGMA code. The contribution of SNe Ia is not shown in this
    panel.}  \lFig{figure_science_case_OMEGA_SYGMA}
\end{figure}

Another widgetized notebook provides an interface for Stellar Yields
for Galactic Modelling Applications \citep[SYGMA,][]{Ritter:2017uy},
which allows to generate simple stellar population models and retrieve
chemical yields among other properties in table formats that can be
used as building blocks for GCE models or hydrodynamic simulations of
galaxy evolution. The SYGMA section of WENDI also contains a
  folder with
  notebooks\footnote{\url{https://github.com/NuGrid/NuPyCEE/tree/master/DOC/Papers/SYGMA_paper}}
  that run the SYGMA code and generate all plots shown in the SYGMA
  code paper \citet{Ritter:2017uy}. This serves as an example how
  \cyberhs\ can be a tool in support of the goal of \emph{reproducible
    science}, by providing not only access to data and code, but also
  to the capability to execute the analysis on the data in a
  controlled and specified environment.

A technical detail concerns how the widget notebooks are launched. The
\cyberhs\ configuration allows automatic self-start up that hides Python
code cells, creating a relatively polished final experience. For this
to work, the widget notebooks have to be designated as
\textit{trusted}. Jupyter references a database in a \textit{notebook
  signatures database} file which contains instances of the notebooks
that are configured to be trusted. In order for a notebook to be
trusted, the raw \code{.ipynb} notebook file must exactly match the
file used to sign said notebook in the database. This system
implemented by {JupyterHub} is rather sensitive to minute changes in the
notebook, and does not easily incorporate notebook updates and
preserve the trusted state of the previously signed notebook. In order
to ensure flexibility of the application hubs with trusted notebooks
in their respective singleuser images, even after a notebook has been
updated, a bash script is used to trust the notebooks in the state
that they exist when staged. This ensures that even if a notebook has
changed remotely prior to building a singleuser application hub with a
reference to an outdated signatures database, the database will be
updated automatically. This script contains a series of paths to the
notebooks in the singleuser environment that the user requires to be
trusted. When its run, it \textit{signs} the \textit{notebook
  signatures database} file in the singleuser notebook directory.

WENDI is provided by the \code{wendihub} docker image that can be
found on docker hub (\code{cyberhubs/wendihub}) as well as in the
astrohubs GitHub
repository\footnote{\url{https://github.com/cyberlaboratories/astrohubs}}.

\subsubsection{NuGrid / \mesa\ experts}
The third of the above cases adds the requirement to install and run
multi-core, parallel simulations. The NuGrid collaboration uses the
\mesa\ code that uses \code{OpenMP} to
provide shared memory parallelism scaling to up to $\approx 10$
cores. Like many sophisticated simulation codes \mesa\ relies on a
significant number of dependencies. The installation processes has
been significantly eased with the
\code{MESA-SDK}.
Still, installation can be a challenge, especially for inexperienced
users. The NuGrid code \code{NuPPN} for single-, multi-zone and tracer
particle processing adopts
\code{MPI} parallelism and exhibits
good strong scaling to $\approx 50$ cores for typical 1D multi-zone
problems. Both applications are compiled with the
\code{gfortran} compiler,
and require \code{hdf} and \code{se} libraries, as well as numerical
libraries, such as
\code{SuperLU}
and \code{OpenBLAS}.  The
singleuser application \code{mesahub} combines the compilers,
libraries and environment variable settings needed to install and run
MESA and NuGrid simulation codes, and probably several other codes
with the same requirements. The presently latest \code{mesahub} docker
image (version 0.9.5) runs
the NuGrid codes \code{NuPPN/mppnp} for parallel multi-zone
simulations, and \code{NuPPN/ppn} for single-zone simulations, and has
been tested for MESA versions 8118, 8845, and 9331. It should be
straight-forward to update this application to accommodate newer as
well as older MESA versions.  The resources of the entire host machine
can be accessed by each user through their application docker
container. We are currently running \cyberhs\ on several servers,
including one instance on a virtual workstation with 16 cores and
120GB memory which allows several concurrent \mesa\ runs as well as
using all 16 cores for \code{NuPPN} multi-zone simulations.

In addition, this application includes a wide
range of Python packages for data analysis, including NuGrid's
\code{NuGridPy} tool box. The application also includes common
command-line editors and a complete \LaTeX\ installation that allows
manuscript generation, complemented with the browser's pdf
viewer. Jupyter extensions that allow easy generation of interactive
slide shows from notebooks for presentations are included. It is
therefore possible to perform all steps needed for a research project
just inside the \code{mesahub} application.

\subsection{PPMstar}
Another application that we want to highlight is the \code{PPMstarhub}
which provides analytic access to stellar hydrodynamics simulations
\citep{herwig:14,Woodward:2013uf,Jones:2017kc}. The challenges in this
case are a combination of very large data sizes and the benefits from
using legacy software in a shared environment. This section starts
with a historical perspective that provides context for the current
development described in this paper.

\subsubsection{Data representation strategies for 3D hydrodynamics simulations with PPMstar}
Over many years, the team at the University of Minnesota's Laboratory
for Computational Science \& Engineering (LCSE) has developed a series
of tools to deal with the voluminous data that is generated by
collections of large 3-D fluid dynamics simulations.  The LCSE was
formed in 1995, but the activity began in 1985 as a result of the
University of Minnesota's purchase that year of the largest and most
powerful supercomputer then available, the Cray-2.  Only 3 of these
machines existed in the world at that time.  This purchase, coupled
with the rarity at the time of academic researchers with simulation
codes capable of exploiting this machine, produced an unprecedented
opportunity. The university had purchased the machine, but not a data
storage system.  An early way around that problem for our research
team was to take advantage of a holiday sale of disk drives by Control
Data, and later the purchase of a tape drive and a small computer to
drive it.  With this experience, a long tradition of ever greater data
compression and development of data analysis and visualization tools
began.  At the time, there were no data file format standards, nor
were there tools, aside from programs one could write oneself, to read
such files. The result of this combination of circumstances was that,
in the LCSE, we developed our own very powerful tools and techniques
to analyze and visualize 3-D simulation data.  As the field has grown,
other groups have taken it upon themselves to produce, enhance, and
maintain such tools for community use, which is a full-time activity
that we chose not to engage in. The infrastructure described in this
article has provided a framework in which we can embed our LCSE tools,
giving them an interface that can be quickly understood and utilized
by others through a Web browser.  Python is the glue that connects our
utilities to the framework and to external users.  All this can now
make our simulation data available to a community of interested
parties around the globe.

Simulations in three dimensions pose special challenges to the
understanding of the computational results.  LCSE did not embark on
3-D simulation until we heard from a Pixar representative in the mid
1980's about the invention of volume rendering.  Once we saw on a
workstation screen the rotating image of the volume rendered water rat
that constituted the first demo of the technique, it was obvious to us
that this was the solution to the data exploration problem for our
fluid dynamics domain.  Soon thereafter, we had our first volume
rendering program, written by David Porter, running on the Cray-2.
This new visualization technique \citep{porter:89,ofelt:89} prompted
the conversion of our 2-D hydrodynamics code to do 3-D
simulations. Initially, we tried to save as much of our simulation
data as we possibly could, because 3-D fluid dynamics simulation was
so new that we thought that we could not possibly predict what
representations of the data we might later wish to make. We quickly
discovered that we could compress saved data down from 64 to 16 bits
per number, saving a factor of 4 in data volume.  Even so, the data
volume was enormous.  Decades of experience with visualizing and
analyzing 3-D simulation data
\citep{woodward:92a,woodward:92b,woodward:93,tucker:93} have led us to
the approach described below that is connected to the Python-based
\cyberhs\ framework described in this article.

Our simulations fall into fairly simple categories, such as
homogeneous turbulence, stellar convection in slab or spherical
geometry, or detailed studies of multifluid interface instability
growth in slab or spherical geometry.  In each category, we do many
simulations that all have common features.  This has meant that after
the first few simulations are completed, we have a very clear idea of
what data we want to preserve and what visualizations we want to make
of the flows in any new category.  We have also developed highly
robust nonlinear maps from the real line, or the positive real line,
to the interval from 0 to 255.  Each such map is determined by an
initial functional transformation, such as a logarithm, for example,
followed by the standard nonlinear map given by the values of just 2
constant parameters.  Such mappings of simulation data to the 256
color levels used in volume rendering can work over all the runs in a
single category of simulations without any modification.  This is not
only tremendously convenient, but it also allows direct and meaningful
visual comparisons of data from different runs.  It turns out that
certain color and opacity maps can work well for a particular
simulation variable, such as vorticity magnitude, over an entire
category of runs without any modification.  These to some degree
unexpected findings have profound consequences for data compression.

\begin{figure*}[hbt]
  \centering
  \includegraphics[width=0.93\textwidth]{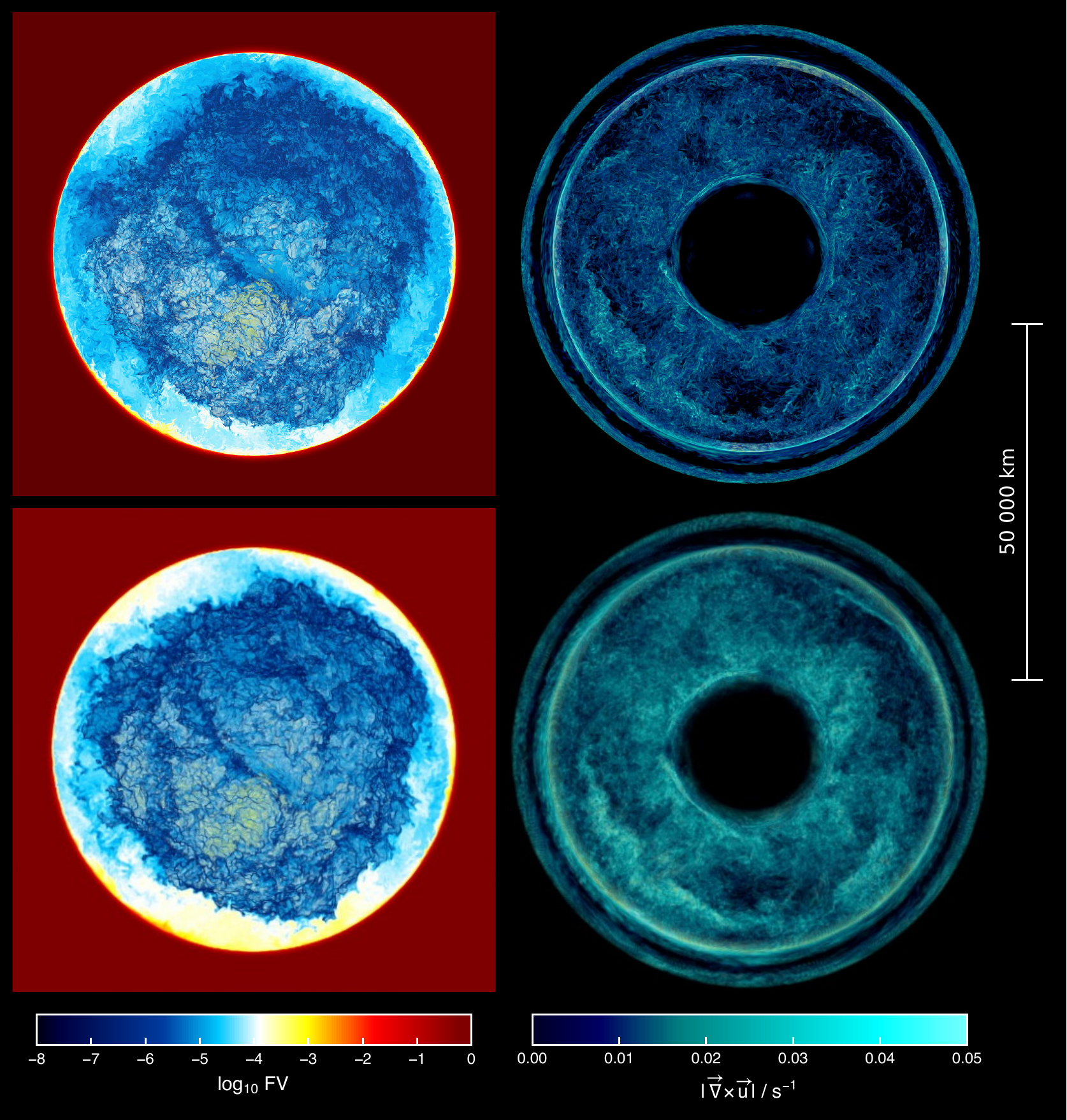}
  \caption{\lFig{briquette_example} Volume renderings of the volume
    mixing fraction (left) of hydrogen-rich material pulled into the
    \pdcz\ in a $2\Msun$, low metallicity star and of the magnitude of
    vorticity (right) in this star.  In the left column the back
    hemisphere of the star's central region is shown, while the
    vorticity images on the right render a thin slice through the 3D
    $4\pi$ simulation domain. The volume renderings at the top are
    made using the full-resolution simulation data, while those at the
    bottom use the briquette-averaged data (see text for details). }
\end{figure*}
The robustness of nonlinear mappings from the real line to color
levels allows us to have our codes dump out only a single byte per
grid cell per variable field.  This is an enormous savings in data
volume.  It can only be helpful if one can know before doing the
simulation which variables are useful for visual exploration of the
simulation results.  One can save even more data volume if one knows
which views of such variables one wants to preserve.  Such images can
then be compressed further by standard image file formats.  Making
this data compression also requires that one know the color and
opacity mapping one wishes to use.  After a run is completed, these
images can now be animated using standard movie making software, such
as
\code{mencoder},
that have replaced our own movie animation software.  However, to save
only images and not the much more voluminous raw voxel data, one must
build the volume renderer into the simulation code.  We have done this
using the \code{srend} software package  \citep{wetherbee:15}.  In this way
rendering with \code{srend} happens right in the code rather than as a
separate activity as part of a complex workflow, which has many
benefits. The full-resolution voxel data need never be written to disk
at all, although, for now, we still write this out just in case.  This
new capability replaces the previous workflow involving the LCSE
\code{HVR} volume renderer that required slow and difficult data
format conversion, GPUs, and special software libraries to run.
\code{srend} is written in Fortran, is compiled along with the
simulation code, and it has no dependencies upon software
libraries. Using either the new \code{srend} capability or the
previous \code{HVR} volume renderer, our strategy is to create default
image views of a pre-defined set of variables for all dumps. These
image libraries are available through the \cyberh.

Even if we were to continue to save full-resolution voxel data, these
data sets are very clumsy to work with due to their sheer size of 45GB
per dump depending on how many variables are saved. At the same time,
a flexible capability to make any visualization of any variable after
the simulation is highly desirable. We accomplish this by performing
an additional data compression.  This final data compression has
evolved from our use of simulation data to validate and develop
statistical models of turbulence \citep{woodward:06a}.

Turbulence closure models deal in averages.  They tend to be based on
comparisons of averages of products and the products of the
corresponding averages.  To work with such models in the early 2000s,
we used averages of our simulation data taken over cubes 32 cells on a
side.  Our filter represented the behavior of a quantity inside the
filter cube by a quadratic form determined by the 10 lowest order
moments of the quantity.  This filtering technique was derived from
our work with the PPB advection scheme
\citep{woodward:86,Woodward:2013uf}, which also works with
the 10 lowest moments. To be able to construct such filtered
representations using a moving filter volume, from our simulations we
saved averages of many different variables taken over cubes 4 grid
cells on a side. We saved these with 16-bit precision, after first
passing them through our robust nonlinear maps.

Our present simulation codes all now work with fundamental data
structures consisting of briquettes of 4 grid cells on a side. The
problem domain is subdivided into regions, which are rectangular
solids, and the regions are subdivided into grid bricks, which are
smaller rectangular solids.  Each grid brick is a brick of briquettes,
augmented all around by a single layer of ghost briquettes from the 26
neighbor bricks.
 
For the storage required to save just one byte of data from each grid
cell in our simulation at any dump time level, we can instead save,
with 16-bit precision, 32 variable averages for each grid briquette of
64 cells.  32 variables is so many that we can save several other
quantities which require differentiating the simulation data and are
useful for model building, in addition to storing variables that we
nearly always look at. We include, for example, the magnitude of the
vorticity, the divergence of the velocity, and both volume-weighted
and mass-weighted averages of the velocity components.  The idea is
that from the 32 quantities one could derive almost anything one may
need when analyzing the data.

Data cubes with the 32 variables representing each of the 216 or 512
regions of a large simulation are saved directly by the code as it
runs in separate disk files.  In this way the analytic tools have
immediate and targeted access to any desired region. In practice, it
has been somewhat a surprise how useful the briquette-averaged data
actually is. While a fine grid is needed in order to advance the
solution in time with high accuracy, this same fine grid is not needed
to represent the solution.  Volume-rendering from full-resolution
images of the mixing fraction and the vorticity are compared with
renderings based on the briquette-averaged lower-resolution data sets
in \Fig{briquette_example}. The full-resolution data consists of a
single-byte voxel value at each cell of the uniform Cartesian grid to
represent each variable we wish to volume render. For the mixing
fraction full-resolution means double resolution, i.e.\ in this case
$3072^3$ voxels, due to the sub-grid resolving power of the
higher-order PPB advection scheme \citep[see appendix
  of][]{Woodward:2013uf}. The high-resolution rendering of the
vorticity and other quantities is based on the grid-resolution
data. The lower-resolution renderings of both mixing fraction and
vorticity are based on data cubes with four times fewer grid points
along each axis. The $384^3$ data cubes contain averages of
$4\times4\times4$-cell briquettes, which for the mixing fraction
represent $8\times8\times8$ significant data values. The images shown
on the top row of \Fig{briquette_example} use the full-resolution
data, and of course give the best representation.  The most important
role of renderings and 3D visualisations is to allow a qualitative
assessment of the flow, that will ultimately guide quantitative model
building.  Even renderings based on the 64-times less voluminous
briquette-averaged data shown on the bottom row still expose most of
the key features of the flow.
   
\Fig{briquette_example} shows the far hemisphere of the He-shell flash
convection or \pdcz\ (similar to those shown in
\Fig{WENDI_kip_contour} and \Fig{WENDI_iso_abu}, see
\Sect{nugrid_application}) in a low-metallicity AGB star.  The inner
core, including the He shell and a stable layer above that contains
H-rich, unprocessed envelope material, is included in the simulation
volume. The gas is confined by gravity and also by a reflecting
boundary sphere at a radius of \unit{33.5}{Mm}.  The star is shown in
the midst of a global oscillation of shell hydrogen ingestion
\citep[GOSH,][]{herwig:14}.  H-rich gas is entrained into the
\pdcz\ from just above the top of the convection zone, at a radius of
about \unit{28}{Mm}.  Waves of combustion involving this mixed-in H
and the \isotope{}{12}{C} are propagating around the outer portion of
the \pdcz.  The convection zone has been formed by the helium shell
flash, with helium burning located at the bottom of the convection
zone, around a radius of \unit{13}{Mm}.  The combustion of entrained
gas at radii around \unit{17}{Mm} drives strong local updrafts, which
greatly enhance convective boundary mixing as the combustion waves
propagate.  This is of course best seen in a movie animation.

In \Fig{briquette_example}, two counter-propagating wave fronts have
recently collided in the region of the lower-left, and a clearly
visible puff of entrained gas has been forced downward there, helping
to form what will soon become a shell of H-enriched gas floating near
the top of the convection zone.  In the images at the left the gas of
the star's carbon-oxygen core as well as the helium-and-carbon mixture
of the convection zone are rendered as transparent.  The color map
represents only the H-rich gas component, that initially is confined
to the stable layers above the convective boundary. At the point of
the simulation shown, some amount of the H-rich fluid has been
entrained into the \pdcz. From the assumed viewing perspective one
sees the lowest H-concentrations first.  Where these, rendered dark
blue, are sufficiently low, one sees through them into the more highly
enriched gas.  At this stage, the lower surface of the newly formed
shell of H-enriched gas is fairly easily deformed as rising plumes of
hotter, more buoyant gas tend to force it aside as they decelerate and
ultimately reverse their upward flow.  Ridges of dark blue in these
images show the lanes that separate neighboring upwellings.  These
ridges of H-enriched gas are descending, pulling the gas from above
the convection zone deeper into the hot region below, where the
hydrogen will burn and drive new waves of combustion. All of these key
features are clearly discernible in the low-resolution renderings
shown in the bottom row of \Fig{briquette_example}.

Volume renderings of the vorticity magnitude (right column,
\Fig{briquette_example}) reveal a thin, highly turbulent region at the
front of the descending puff of entrained gas at the lower left. A
typical, unstable shear layer induced by boundary-layer separation
\citep{Woodward:2013uf} can be seen in the upper right quadrant. These
features have been identified as an important component of the 3D
entrainment mechanism at boundaries that are Kelvin-Helmholtz stable
according to the 1D radial stratification.  The images based on the
briquette-averaged data gives a fuzzy, slightly out-of-focus
impression. But they still clearly reveal these key features of the
flow, and, when such images are animated, their dynamics.

\begin{figure}[hbt]
  \includegraphics[width=\columnwidth]{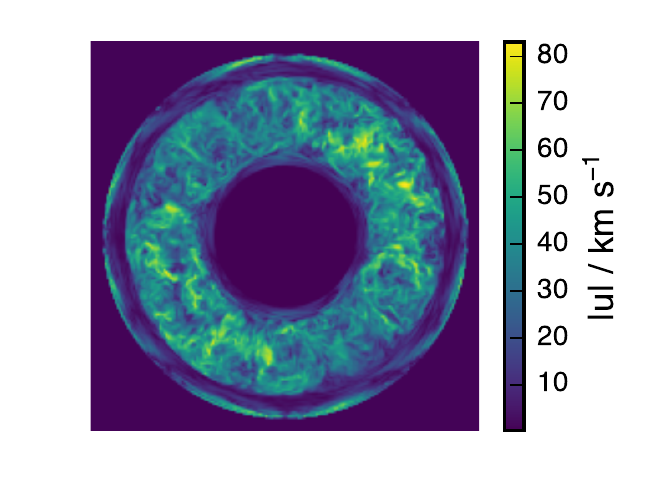}
  \includegraphics[width=\columnwidth]{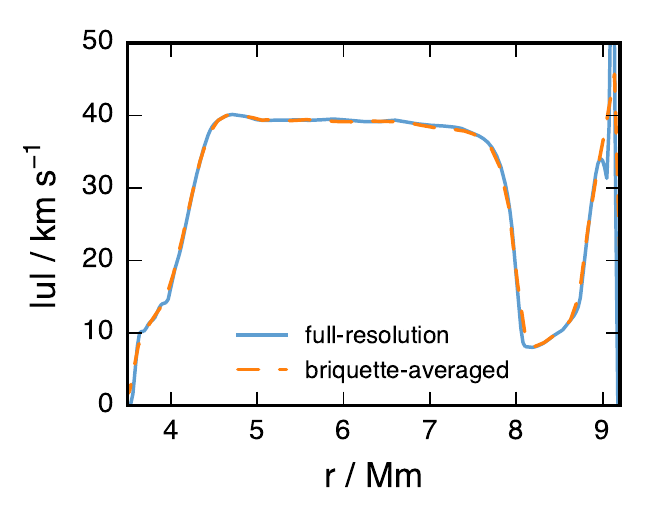}
  \includegraphics[width=\columnwidth]{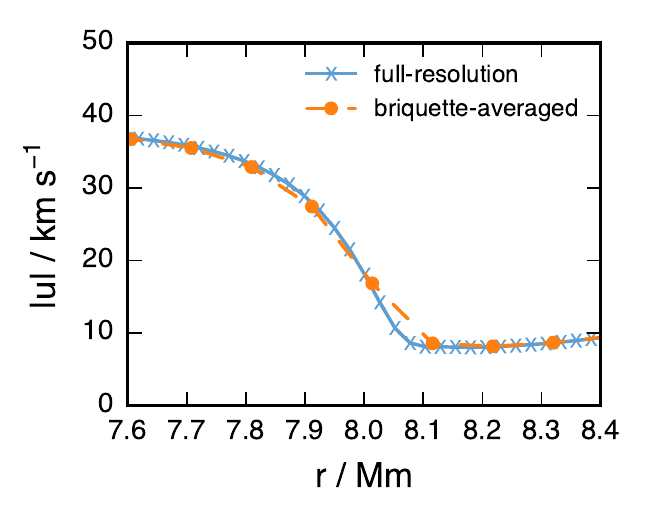}
  \caption{A center-plane slice showing the convective speed from the
    $768^3$-grid O-shell convection simulation D1 presented by
    \citet{Jones:2017kc}, based on briquette-averaged 3D data (top
    panel), as well as spherically averaged profiles based on
    full-resolution data and on briquette-averaged
    data. \lFig{D15_0200} }
\end{figure}
Another more quantitative example of how the briquette-averaged data
can be used is shown in \Fig{D15_0200}. Even at a moderately-sized
grid of $768^3$ the down-sampled data provides a good initial
impression of the overall structure of the flow at a computational-
and data-related cost that can be easily accommodated in the analysis
scenario of the \code{PPMstarhub}. For the overall speed profile the
low- and high-resolution data representations can hardly be
discerned. Even when zooming in to the upper convective boundary the
low resolution data provides meaningful exploratory information.
   
\subsubsection{The PPMstar application hub}
We have used the \cyberhs\ technology to build the \code{PPMstarhub}
application which is addressing several issues. The \code{PPMstar}
simulations of stellar convection are performed on the Blue Waters
computing system at the NCSA center on as many as 400,000 cores. The
resulting data sets are in their own way unique, similar to
astronomical surveys. Although our team is exploiting them for their
primary scientific purpose, there are potentially a number of
additional questions that these data sets could answer. Sharing the
raw data by just making it available for download is impractical due
to the size of the data, as well as the specialized analytic tools
that are needed to access and explore the data.  \cyberhs\ allows us to
expose to interested users these simulation data sets together with
our specialized software stack for analysis.

Over the years, a full range of tools have been developed at the LCSE
that exploit the briquette data sets, both for visualization as well
as for further analysis that would be involved, for example, in model
building. As mentioned above, such models could be turbulence models,
or mixing models, to ultimately be deployed in 1D stellar evolution
models. These tools are now available along with access to several of
our published data sets to interested users through
PPMstarHub\footnote{\url{https://hickory.lcse.umn.edu}}.

In addition, we have developed additional new Python-based analytic
tools that work with the briquette data as well as with single and
multiple radial profile data. These, along with collections of example
notebooks are available on the PPMstar GitHub
repository\footnote{\url{https://github.com/PPMstar}}. Specifically,
the examples include notebooks that contain the analysis of all plots
shown in our recent study on stellar hydrodynamics of O-shell
convection in massive stars \citep{Jones:2017kc}, as well as the
notebooks of our project of simulations of low-Z AGB H-ingestion into
a thermal-pulse He-shell flash (Woodward et al., in prep). All of
these example notebooks are staged on our PPMstarHub server where the
necessary data sets are staged as well, so that interested users can
follow our data analysis. This is an example of how \cyberhs\ can play
an essential role in making scientific analysis of raw simulation or
observational data more transparent and accessible, and the process of
data analysis reproducible.

\subsection{How to add new applications}
The previous sections have described two use cases that have guided
the requirements for \cyberhs. Adopting \cyberhs\ for a different
application would start either with the \code{corehub} application, or
with one of the already existing applications. One would modify the
requirements files that specify the Python and Linux software stack,
and, following the examples provided, add any custom software and
tools required.  For example, we have built a basic
\cyberhs\ application image for machine learning (\code{mlhub}) and
intend to evolve this into a StarNet application for users and
developers. StarNet is an application of deep neural networks for the
analysis of stellar spectra and especially abundance determination
\citep{Fabbro:2017vx}. If one builds a StarNet hub on top of WENDI hub
users could perform combined analysis of stellar abundance
determination and interpretation of these abundances using, for
example, the NuPyCEE tools.

We have also created targeted application for teaching specific
courses, such as the second-year computational physics and math
course\footnote{\url{https://github.com/cyberlaboratories/teachinghubs},
  \url{https://hub.docker.com/r/cyberhubs/mp248}} at the University of
Victoria, that we are currently teaching with 90 students on the
\code{mp248} application that can be optionally launched on the server
that also offers the WENDI application.

\section{Conclusions}
\lSect{discussion} Leveraging dockers, jupyterhub, jupyterlab and
jupyter notebook, we designed, implemented and deployed the
\cyberhs\ system. It provides collaborations and research groups with a
common collaboration platform in which data, analytic tools,
processing capacity as well as different levels of user interactions
(Python or bash notebooks, terminals, GUI/widget
notebooks). \cyberhs\ adopts a simple, flexible and effective access and
authorization model.

The system is easy to deploy, to customize, and is already in
production. By pulling a docker image, cloning a GitHub repository and
specifying a few environment variables, administrators can launch VREs
for their users.  Existing core or more advanced application hub
images can be customized to suit specialized needs. In addition to the
basic corehub, our specialized hubs are in production and used for
collaborations such as NuGrid and PPMstar, as well as in the class
room teaching classes with dozens of students.

\subsection{Limitations and future development}
As with any multiuser platform, the \cyberhs\ require designated
personnel to administrate and maintain, though the deployment of our
hubs is straight-forward.  The fact that we are dealing with leading
edge technologies makes the system susceptible to major changes at any
time. We have protected \cyberhs\ to some degree by enforcing
version-locking of each included Python and Linux software package.
Although we have frozen the pip and apt requirements by specifying for
each component the version to be used at build time to avoid package
incompatibility issues, security updates of any package may require
the users to update the system.  Another limitation of the system is
that it does not scale with the number of users, and does not offer
resource allocation or scheduling capabilities.  As the number of
users increases, a \cyberh\ may run out of resources and larger servers
may become necessary.  Therefore, we are exploring Docker Swarm and
Kubernetes to scale the resources and schedule containers into
distributed resources.  Volume selection is also an important feature
that we plan to add.  Depending on the credentials of a user, we are
interested in ensuring that the user has access to special/private
volumes.  We also plan to add \code{letsencrypt} capability to our
hubs so that administrators are freed from dealing with SSLs directly.

We invite those who are creating new application hub images to share
these through adding the build files to the astrohhubs GitHub
repository and to submit such images to be pushed to the
\cyberhs\ Docker Hub organisation \cyberhs.

\acknowledgements The \cyberhs\ project is building on a previous
Canarie funded CANFAR project \emph{Software-as-a-service for Big Data
Analytics} in which the first version of WENDI was built with
pre-\code{JupyterHub} tools. Further funding was provided by NSERC USRA,
NSERC Discovery, EcoCanada and the National Science Foundation (NSF)
under Grant No. PHY-1430152 (JINA Center for the Evolution of the
Elements). Previous undergraduate students in the Coop program of the
Department of Physics and Astronmy at the University of Victoria who
have directly or indirectly contributed are William Hillary and Daniel
Monti, who developed the initial versions of the \code{NuGridPy}
software. Luke Siemens has made significant contributions to an
initial version of the new, and more general version of WENDI based on
\code{JupyterHub}.  The data sets and software tools in NuGrid's WENDI
\cyberh\ were developed by members of the NuGrid collaboration
(http://www.nugridstars.org). The motivation for this paper described
in the introduction was previously expressed, in part, in CANFAR's CFI
proposal "Astronomy Cyber-laboratories Platform", PI Falk Herwig,
submitted in October 2017. We also acknowledge support for our large
simulations on the Blue Waters machine at NCSA with \code{PPMstar}
from NSF PRAC awards 1515792 and 1713200, and support for work at
Minnesota on these simulations and construction of means to serve and
share the data from NSF CDS\&E grant 1413548.

\software{
  \code{Juypter notebook} \url{http://jupyter.org},
  \code{Jupyterlab} \url{https://github.com/jupyterlab},
  \code{VOspace} \url{http://www.canfar.net/en/docs/storage},
  \code{vos} \url{https://pypi.python.org/pypi/vos},
  \code{VirtualBox} \url{https://www.virtualbox.org},
  \code{JupyterHub} \url{https://jupyterhub.readthedocs.io/en/latest/},
  \code{ipywidgets} \url{https://ipywidgets.readthedocs.io},
  \code{NuPyCEE} \url{http://nugrid.github.io/NuPyCEE},
  \code{NuGridSetExplorer} \url{https://github.com/NuGrid/WENDI},
  \code{hdf5} \url{https://www.hdfgroup.org},
  \code{Cyberlaboratories} cyberhubs \url{https://github.com/cyberlaboratories/cyberhubs},
  \code{Cyberlaboratories} astrohubs \url{https://github.com/cyberlaboratories/astrohubs},
  \code{Cyberhubs Docker repository} \url{https://hub.docker.com/u/cyberhubs},
  \code{Docker} \url{https://www.docker.com},
  \code{NOAO data lab} \url{http://datalab.noao.edu},
  \code{ansible} \url{https://www.ansible.com},
  \code{puppet} \url{https://puppet.com},
  \code{mesa\_h5} \url{https://github.com/NuGrid/mesa\_h5},
  \code{Python} \url{https://www.python.org},
  \mesa\ \url{http://mesa.sourceforge.net},
  \code{WENDI} \url{http://wendi.nugridstars.org},
  \code{OpenMP} \url{http://www.openmp.org},
  \code{MESA-SDK} \url{http://www.astro.wisc.edu/~townsend/static.php?ref=mesasdk},
  \code{MPI} \url{https://www.open-mpi.org},
  \code{gfortran} \url{https://gcc.gnu.org/fortran},
  \code{SuperLU} \url{http://crd-legacy.lbl.gov/~xiaoye/SuperLU},
  \code{OpenBLAS} \url{http://www.openblas.net},
  \code{mencoder} \url{http://www.mplayerhq.hu}  
}


\begin{thebibliography}{}
\expandafter\ifx\csname natexlab\endcsname\relax\def\natexlab#1{#1}\fi
\providecommand{\url}[1]{\href{#1}{#1}}

\end{thebibliography}


\begin{thebibliography}{}
\expandafter\ifx\csname natexlab\endcsname\relax\def\natexlab#1{#1}\fi
\providecommand{\url}[1]{\href{#1}{#1}}

\bibitem[{{Eggenberger} {et~al.}(2008){Eggenberger}, {Meynet}, {Maeder},
  {Hirschi}, {Charbonnel}, {Talon}, \& {Ekstr{\"o}m}}]{Eggenberger:2008}
{Eggenberger}, P., {Meynet}, G., {Maeder}, A., {et~al.} 2008, \apss, 316, 43

\bibitem[{Fabbro {et~al.}(2017)Fabbro, Venn, O'Briain, Bialek, Kielty,
  Jahandar, \& Monty}]{Fabbro:2017vx}
Fabbro, S., Venn, K., O'Briain, T., {et~al.} 2017, eprint arXiv:1709.09182,
  1709.09182

\bibitem[{{Gallino} {et~al.}(1998){Gallino}, {Arlandini}, {Busso}, {Lugaro},
  {Travaglio}, {Straniero}, {Chieffi}, \& {Limongi}}]{gallino:98}
{Gallino}, R., {Arlandini}, C., {Busso}, M., {et~al.} 1998, \apj, 497, 388

\bibitem[{Herwig(2005)}]{herwig:04c}
Herwig, F. 2005, ARAA, 43, 435

\bibitem[{Herwig(2013)}]{Herwig:2013fc}
---. 2013, in link.springer.com (Dordrecht: Springer Netherlands), 397--445

\bibitem[{{Herwig} {et~al.}(2014){Herwig}, {Woodward}, {Lin}, {Knox}, \&
  {Fryer}}]{herwig:14}
{Herwig}, F., {Woodward}, P.~R., {Lin}, P.-H., {Knox}, M., \& {Fryer}, C. 2014,
  \apjl, 792, L3

\bibitem[{Jones {et~al.}(2017)Jones, Andr{\'a}ssy, Sandalski, Davis, Woodward,
  \& Herwig}]{Jones:2017kc}
Jones, S., Andr{\'a}ssy, R., Sandalski, S., {et~al.} 2017, MNRAS, 465, 2991

\bibitem[{Jones {et~al.}(2014)Jones, Herwig, Siemens, Fabbro, Gaudet,
  Pignatari, \& Trappitsch}]{jones:14}
Jones, S., Herwig, F., Siemens, L., {et~al.} 2014, poster contribution at
  Nuclei in the Cosmos NIC XIII conference, available at
  \url{http://www.nugridstars.org/publications/conference-contributions/nuclei-in-the-cosmos-xiii/NIC_CADC_poster_best.pdf}

\bibitem[{Lodders {et~al.}(2009)Lodders, Palme, \& Gail}]{Lodders:2009er}
Lodders, K., Palme, H., \& Gail, H.~P. 2009, in Solar System (Berlin,
  Heidelberg: Springer Berlin Heidelberg), 712--770

\bibitem[{Ofelt {et~al.}(1989)Ofelt, Porter, Varghese, Schmid, Ruwart, \&
  Woodward}]{ofelt:89}
Ofelt, D., Porter, D., Varghese, T., {et~al.} 1989, PPM Graphics Tools, Tech.
  rep., Minnesota Supercomputer Institute

\bibitem[{Paxton {et~al.}(2010)Paxton, Bildsten, Dotter, Herwig, Lesaffre, \&
  Timmes}]{Paxton2011}
Paxton, B., Bildsten, L., Dotter, A., {et~al.} 2010, \apjs, 192, 3

\bibitem[{Paxton {et~al.}(2013)Paxton, Cantiello, Arras, Bildsten, Brown,
  Dotter, Mankovich, Montgomery, Stello, Timmes, \& Townsend}]{Paxton:2013km}
Paxton, B., Cantiello, M., Arras, P., {et~al.} 2013, ASTROPHYS J SUPPL S, 208,
  4

\bibitem[{Paxton {et~al.}(2015)Paxton, Marchant, Schwab, Bauer, Bildsten,
  Cantiello, Dessart, Farmer, Hu, Langer, Townsend, Townsley, \&
  Timmes}]{Paxton:2015iy}
Paxton, B., Marchant, P., Schwab, J., {et~al.} 2015, ASTROPHYS J SUPPL S, 220,
  15

\bibitem[{Pignatari \& Herwig(2012)}]{Pignatari:2012dw}
Pignatari, M., \& Herwig, F. 2012, Nuclear Physics News, 22, 18

\bibitem[{Pignatari {et~al.}(2016)Pignatari, Herwig, Hirschi, Bennett,
  Rockefeller, Fryer, Timmes, Ritter, Heger, Jones, Battino, Dotter,
  Trappitsch, Diehl, Frischknecht, Hungerford, Magkotsios, Travaglio, \&
  Young}]{Pignatari:2016er}
Pignatari, M., Herwig, F., Hirschi, R., {et~al.} 2016, ASTROPHYS J SUPPL S,
  225, 24

\bibitem[{Porter \& Woodward(1989)}]{porter:89}
Porter, D.~H., \& Woodward, P.~R. 1989, in ACM SIGGRAPH Video Review, Vol.~44,
  Volume Visualization State of the Art, ed. L.~Herr, 10-minute video segment:
  Simulations of Compressible Convection with PPM

\bibitem[{{Ritter} \& {C{\^o}t{\'e}}(2016)}]{2016ascl.soft10015R}
{Ritter}, C., \& {C{\^o}t{\'e}}, B. 2016, {NuPyCEE: NuGrid Python Chemical
  Evolution Environment},  Astrophysics Source Code Library, ascl:1610.015

\bibitem[{Ritter {et~al.}(2017{\natexlab{a}})Ritter, C{\^o}t{\'e}, Herwig,
  Navarro, \& Fryer}]{Ritter:2017uy}
Ritter, C., C{\^o}t{\'e}, B., Herwig, F., Navarro, J.~F., \& Fryer, C.
  2017{\natexlab{a}}, ApJS submitted, 1711.09172v1

\bibitem[{Ritter {et~al.}(2017{\natexlab{b}})Ritter, Herwig, Jones, Pignatari,
  Fryer, \& Hirschi}]{Ritter:2017wy}
Ritter, C., Herwig, F., Jones, S., {et~al.} 2017{\natexlab{b}}, MNRAS,
  submitted, arxiv:1709.08677

\bibitem[{{Thielemann} {et~al.}(1986){Thielemann}, {Nomoto}, \&
  {Yokoi}}]{thielemann:86}
{Thielemann}, F.-K., {Nomoto}, K., \& {Yokoi}, K. 1986, \aap, 158, 17

\bibitem[{Tucker \& Woodward(1993)}]{tucker:93}
Tucker, L., \& Woodward, P.~R. 1993, System Software and Tools for High
  Performance Computing Environments (SIAM), 109--113, {Paul Messina and Thomas
  Sterling, eds.}

\bibitem[{Wetherbee {et~al.}(2015)Wetherbee, Jones, Knox, Sandalski, \&
  Woodward}]{wetherbee:15}
Wetherbee, T., Jones, E., Knox, M., Sandalski, S., \& Woodward, P. 2015, in
  Proceedings of the 2015 XSEDE Conference: Scientific Advancements Enabled by
  Enhanced Cyberinfrastructure (ACM), {Article No.\,35}.
\newblock \url{https://dl.acm.org/citation.cfm?doid=2792745.2792780}

\bibitem[{{Woodward}(1986)}]{woodward:86}
{Woodward}, P.~R. 1986, in Astrophysical Radiation Hydrodynamics, ed. K.-H.
  Winkler \& M.~L. Norman, Reidel, 245--326, online at
  http://www.lcse.umn.edu/PPMlogo

\bibitem[{Woodward(1992{\natexlab{a}})}]{woodward:92a}
Woodward, P.~R. 1992{\natexlab{a}}, Segment of “Art of Science” program
  broadcast nationally,  PBS, Scientific American Frontiers, produced at
  Showcase ’92 exhibit at SIGGRAPH

\bibitem[{Woodward(1992{\natexlab{b}})}]{woodward:92b}
Woodward, P.~R. 1992{\natexlab{b}}, in Proc. Supercomputing Japan, scientific
  Visualization of Complex Fluid Flow, also available as Minnesota
  Supercomputer Institute Research Report UMSI 92/233

\bibitem[{Woodward(1993)}]{woodward:93}
Woodward, P.~R. 1993, in IEEE Computer, Vol.~26

\bibitem[{Woodward {et~al.}(2015)Woodward, Herwig, \& Lin}]{Woodward:2013uf}
Woodward, P.~R., Herwig, F., \& Lin, P.-H. 2015, ApJ, 798, 49

\bibitem[{{Woodward} {et~al.}(2006){Woodward}, {Porter}, {Anderson}, {Fuchs},
  \& {Herwig}}]{woodward:06a}
{Woodward}, P.~R., {Porter}, D.~H., {Anderson}, S., {Fuchs}, T., \& {Herwig},
  F. 2006, Journal of Physics Conference Series, 46, 370

\end{thebibliography}

\end{document}